\documentclass[aps,prc,letterpaper,showpacs,12pt,floatfix,preprint]{revtex4-2}

\usepackage{natbib}
\usepackage{amsmath}
\usepackage{graphicx}
\usepackage{hyperref}
\usepackage[utf8]{inputenc}
\usepackage{color}
\usepackage{empheq}

\usepackage{bm}

\allowdisplaybreaks

\begin{document}

	\title{Electromagnetic Single-nucleon Response \\ involving Polarized Targets}
	
	\author{T. W. Donnelly$^{(1)}$ and Sabine Jeschonnek$^{(2)}$}
	
	\affiliation{\small \sl 
		(1) Center for Theoretical Physics, Department of Physics and Laboratory for Nuclear Science, Massachusetts Institute of Technology, Cambridge, MA 02139 \\
		(2) The Ohio State University, Physics
		Department, Lima, OH 45804		 
	}

	\date{\today}

	\begin{abstract}
		
This work is an extension of our past study focused on a covariant representation of the electromagnetic (EM) current of spin-1/2 Dirac particles, specifically, nucleons. In the past study the EM responses that occur in unpolarized electron scattering from unpolarized nucleons were derived; however, scattering of polarized electrons from polarized nucleons was beyond the scope of that earlier work. Here such extensions are studied in detail. While in other work the EM response has already been developed for the double-polarization scattering problem, that effort was focused on high-energy collider physics. In the present study the formalism is recast into a set of EM response functions that have transparent dependencies on the relevant kinematic variables,  especially on how these behave with respect to the momentum {\bf p} of the (moving) struck, polarized nucleon. The motivation for such a reformulation of the problem is the desire to see a clear path to expansions in $p$ of the EM response for use in devising ``prescriptions for nuclear physics''. Results are provided where comparisons of the full (unexpanded) responses with various approximations that are frequently employed in studies of EM nuclear physics are made, demonstrating that under some circumstances such approximations are reasonable, whereas in other circumstances the expanded results are likely to be invalid. In addition, the EM current operators and approximations to them are discussed in detail.
		
	\end{abstract}

	\maketitle


\section{\protect\bigskip Introduction\label{sec-intro}}

The present study provides extensions to our previous one undertaken in \cite{JandD}. This earlier work was focused on unpolarized single-nucleon electron scattering from unpolarized nucleons and, while the basic formalism was provided for the situations where both the incident electron and target nucleon could be polarized, this subject was felt to be beyond scope of the original project. Here we consider that polarized scenario in some depth. We begin in Sec.~\ref{sec-curr} with a review and some new details on the electromagnetic current of single nucleons, building on the previous developments in \cite{JandD}. Importantly, we explain why caution should be exercised when employing simple implementations of what is termed the ``prescription for nuclear physics'' where one goes from some representation of the single-nucleon electromagnetic current --- usually approximated --- to operators for the current to be used in coordinate space. We explain why the usually so-approximated current operators may work in very specific situations, but not in modern applications, namely, in the so-called ``quasielastic'' regime where they are suspect; such situations are central to modern studies with GeV-scale electron scattering or neutrino reactions.

Section~\ref{sec-curr} is followed in Sec.~\ref{sec-elscatt} by a detailed treatment of the single-nucleon electron scattering response functions, both unpolarized (from \cite{JandD}) and their extensions to polarized responses, the latter being considerably more involved to obtain given the need to address the issue of providing the formalism for handling the target nucleon spin. Section~~\ref{subsec-tensors} begins by summarizing the general nature of the electromagnetic hadronic tensors that underlie all electron scattering, using the formalism in \cite{Donnelly:2023rej,arxivlong} as a basis for those discussions. This is followed in Sec.~\ref{subsec-single} by the basic developments required in deriving both the unpolarized and polarized single-nucleon response functions, to be found in Secs.~\ref{subsec-unpol} and \ref{subsec-pol}, respectively. Section~\ref{subsec-low} contains a brief treatment of what happens in situations where the target momentum is small. And then in Sec.~\ref{sec-num} numerical results are provided to quantify the behavior with kinematics of all of the single-nucleon response functions. Finally, in Sec.\ref{sec-concl} we present our conclusions. 

\section{\protect\bigskip On-Shell Single-Nucleon Electromagnetic Current\label{sec-curr}}

\subsection{\protect\bigskip Basic Kinematics\label{subsec-basic}}

\begin{figure}
	\centering
	\includegraphics[width=16cm]{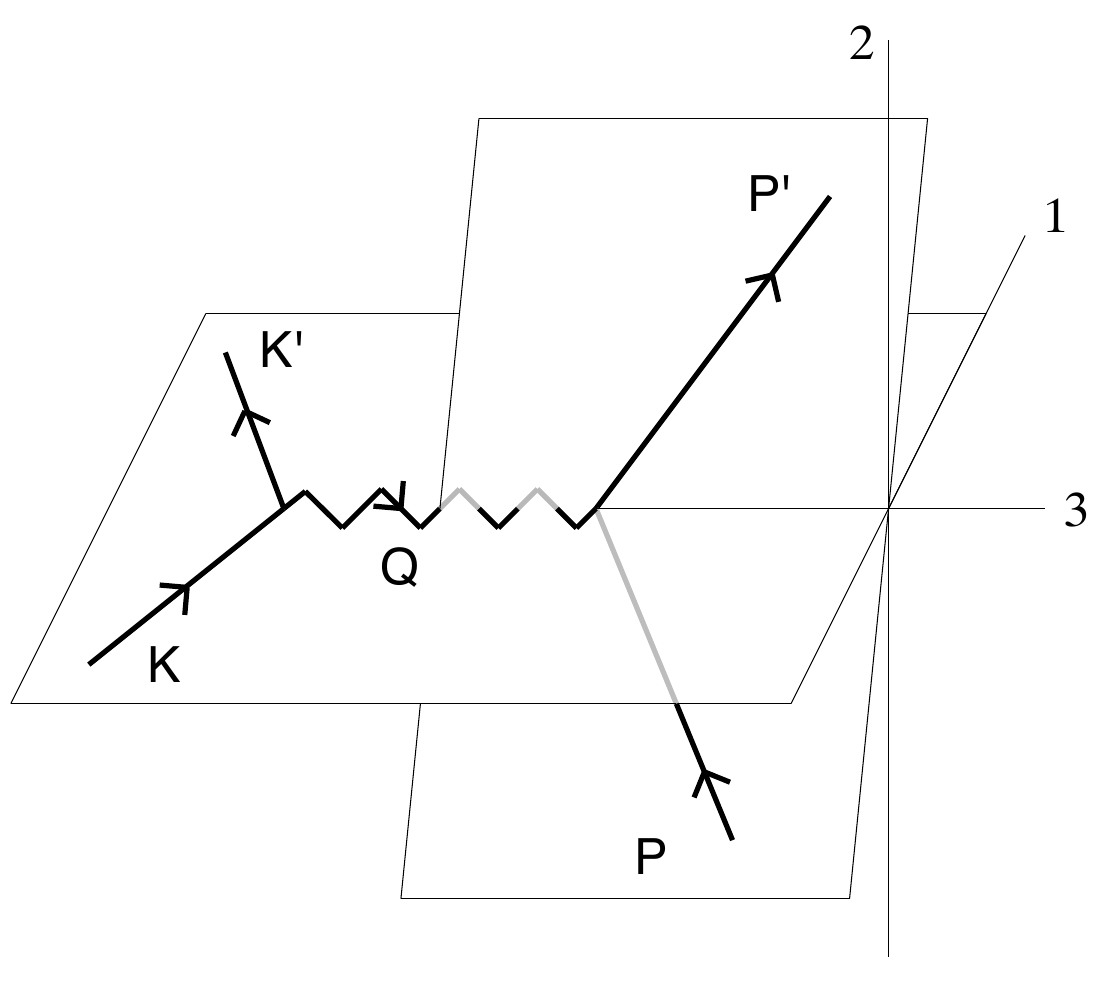} 				
	\caption{General kinematics for single-nucleon electron scattering showing the relevant kinematic variables (see text for details). }
	\label{fig-1}
\end{figure}

In this study we draw on the presentations in \cite{Donnelly:2023rej,arxivlong}, adopting the conventions used in that work (see also Appendix \ref{sec-conventions}). The basic electron scattering reaction is shown schematically in Fig.~\ref{fig-1}, where an incident electron with 4-momentum $K^\mu = (\epsilon, {\mathbf k})$ is scattering to 4-momentum $K'^\mu = (\epsilon', {\mathbf k}')$ and exchanges the 4-momentum $Q^\mu = (\omega, {\mathbf q}) = K^\mu - K'^\mu$ carried by the virtual photon. This implies that $\omega = \epsilon - \epsilon'$ and ${\mathbf q} = {\mathbf k} - {\mathbf k}'$. The virtual photon is absorbed by the target having 4-momentum $P^\mu = (E_p, {\mathbf p})$, going to a final state having 4-momentum $P'^\mu = (E'_p, {\mathbf p}')$. Four-momentum conservation then requires that $Q^\mu = P'^\mu - P^\mu$, which implies that $\omega = E'_p - E_p$ and ${\mathbf q} = {\mathbf p}' - {\mathbf p}$. In the special case of elastic electron scattering from the nucleon of mass $m_N$ that forms the focus of the present study one also has $E_p = \sqrt{m_N^2 + p^2}$ and $E'_p = \sqrt{m_N^2 + p'^2}$. 

We begin by summarizing the various coordinate systems used in the following developments. As in past work the $1 2 3$-system is set up to be convenient for electron scattering from hadronic systems as shown in the figure, where ${\mathbf u}_3$ lies along ${\mathbf q}$, ${\mathbf u}_1$ is in the lepton scattering plane, and ${\mathbf u}_2 = {\mathbf u}_3 \times {\mathbf u}_1$. 
This system is chosen because of the special role played by the virtual photon when invoking the plane-wave Born approximation (single-photon exchange) in that one has ${\mathbf q} = q {\mathbf u}_3$. In the $123$-system one can write ${\mathbf p} = p ( \sin\theta \cos\phi {\mathbf u}_1 + \sin\theta \sin\phi {\mathbf u}_2 + \cos\theta {\mathbf u}_3)$ and ${\mathbf p}' = p' ( \sin\theta' \cos\phi' {\mathbf u}_1 + \sin\theta' \sin\phi' {\mathbf u}_2 + \cos\theta' {\mathbf u}_3)$, with $\phi' = \phi$.

\begin{figure}
	\centering
	\includegraphics[width=12cm]{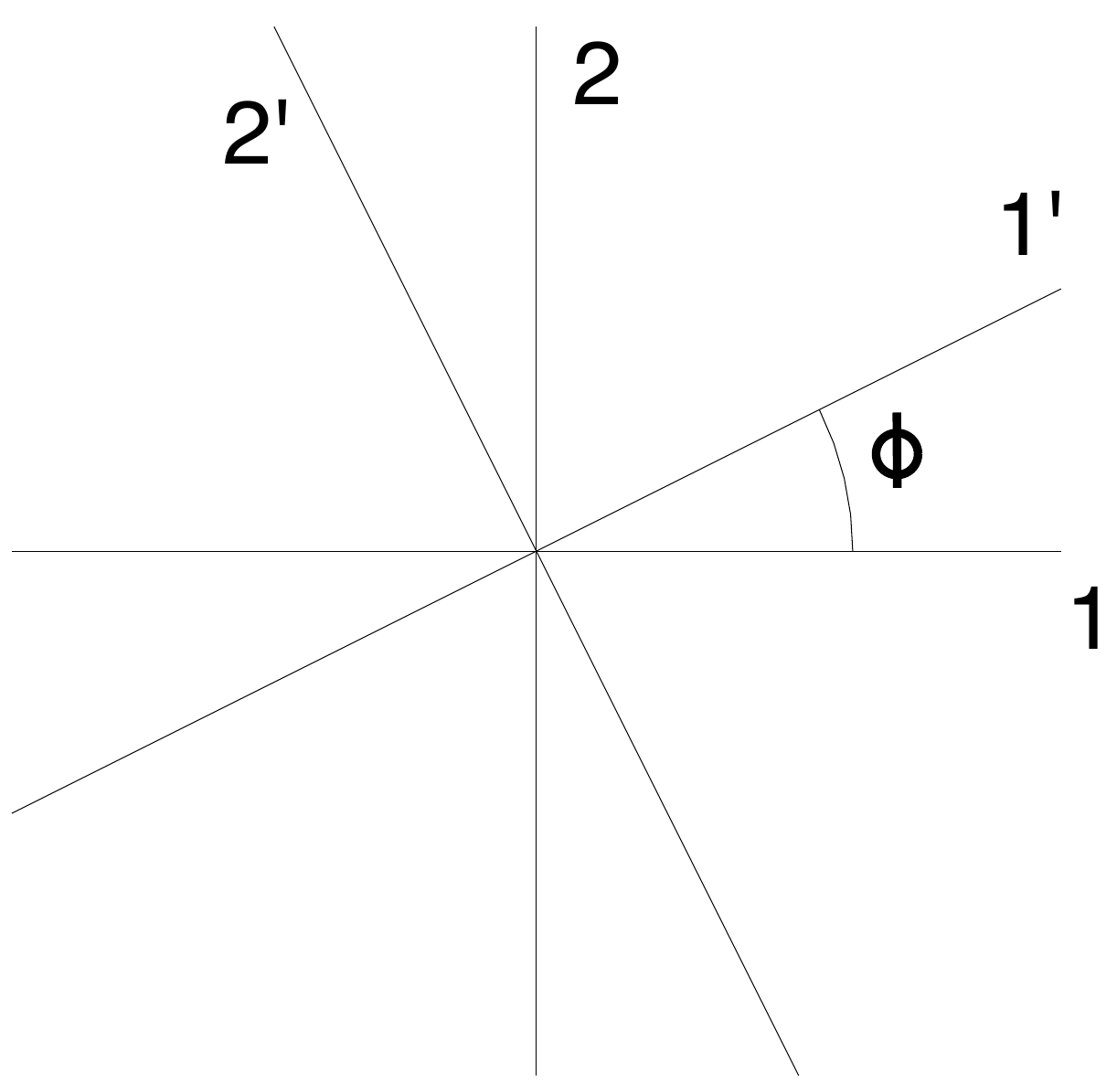} 				
	\caption{Rotation relating the $123$- and $1'2'3'$ systems. Here ${\mathbf u}'_3 = {\mathbf u}_3.$ }
	\label{fig-2}
\end{figure}

It is also convenient to rotate to the $1'2'3'$-system where ${\mathbf u}^{\prime}_1$ lies in the hadron plane as usual (see Fig.~\ref{fig-2}). The two systems are related by the following:
\begin{eqnarray}
{\mathbf u}'_1 &=& \cos\phi {\mathbf u}_1 + \sin\phi {\mathbf u}_2 \nonumber \\ 
{\mathbf u}'_2 &=& -\sin\phi {\mathbf u}_1 + \cos\phi {\mathbf u}_2 \nonumber \\
{\mathbf u}'_3 &=& {\mathbf u}_3 \label{eq-fr-3}
\end{eqnarray}
and
\begin{eqnarray}
{\mathbf u}_1 &=& \cos\phi {\mathbf u}'_1 - \sin\phi {\mathbf u}'_2 \nonumber \\ 
{\mathbf u}_2 &=& \sin\phi {\mathbf u}'_1 + \cos\phi {\mathbf u}'_2 \nonumber \\
{\mathbf u}_3 &=& {\mathbf u}_3 . \label{eq-fr-6}
\end{eqnarray}
In the $1'2'3'$-system one then has ${\mathbf p} = p ( \sin\theta {\mathbf u}'_1 + \cos\theta {\mathbf u}'_3)$ and ${\mathbf p}' = p' ( \sin\theta' {\mathbf u}'_1 + \cos\theta' {\mathbf u}'_3)$. The 4-momentum conservation condition then requires that
\begin{eqnarray}
p' \sin\theta' &=& p \sin\theta \nonumber \\
p' \cos\theta' &=& p \cos\theta + q . \label{eq-4-2} 
\end{eqnarray}

\begin{figure}
	\centering
	\includegraphics[width=12cm]{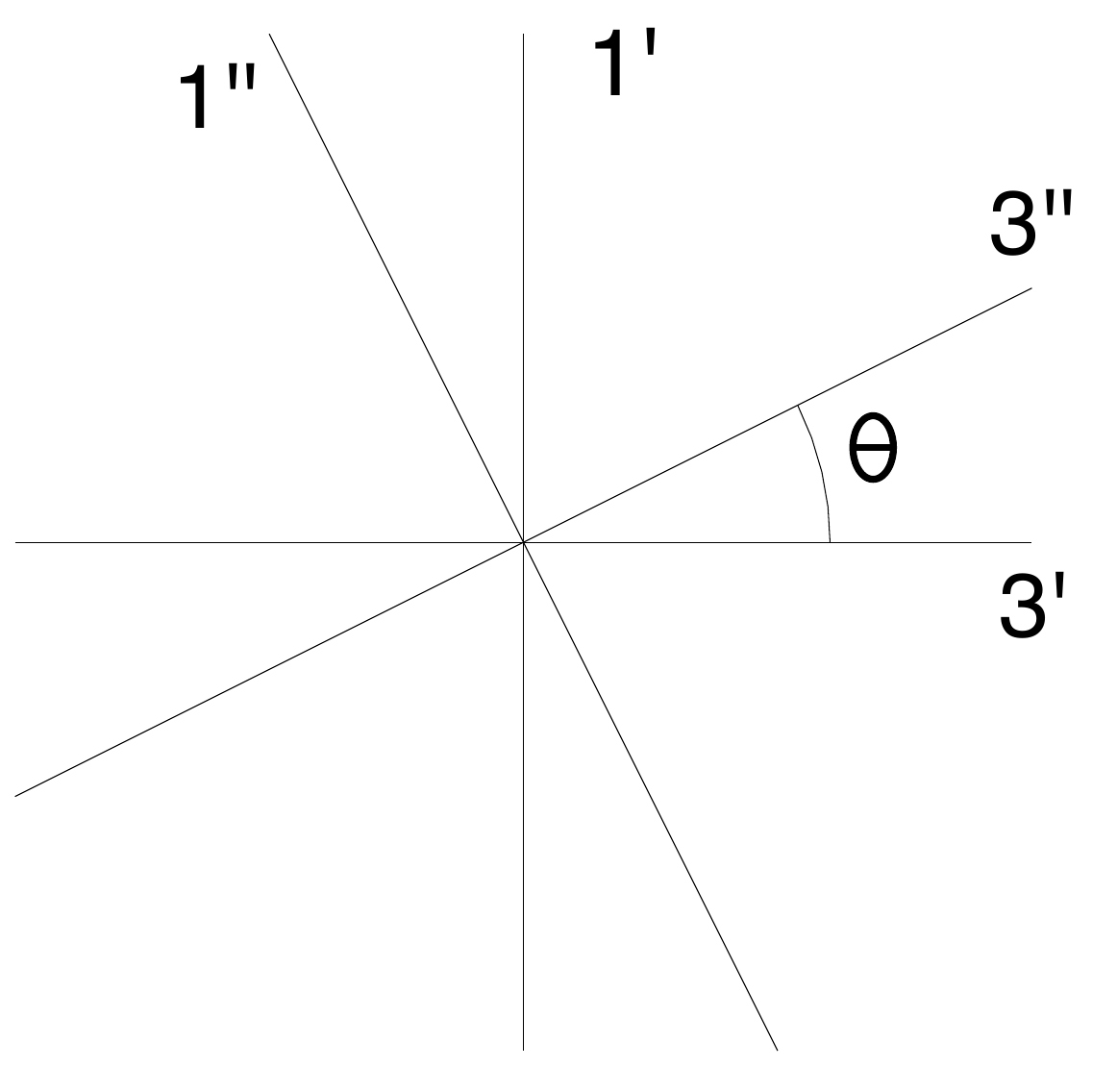} 				
	\caption{Rotation relating the $1'2'3'$- and $1''2''3''$ systems. Here ${\mathbf u}''_2 = {\mathbf u}'_2.$  }
	\label{fig-3}
\end{figure}

Later, when considering the spin content in the problem, it is also useful to employ a coordinate system determined by the direction of the incoming nucleon, {\it i.e.,} using the 3-momentum ${\mathbf p}$. We take this direction to define a unit vector in the direction $3''$ via
\begin{equation}
{\mathbf p} \equiv p {\mathbf u}''_3 , \label{eq-ss-1}
\end{equation}
we take the direction $2''$ to be the same as $2'$,
\begin{equation}
{\mathbf u}''_2 \equiv {\mathbf u}'_2, \label{eq-ss-2}
\end{equation}
and then the remaining unit vector required is given by
\begin{equation}
{\mathbf u}''_1 \equiv {\mathbf u}'_2 \times {\mathbf u}''_3 ; \label{eq-ss-3}
\end{equation}
see Fig.~\ref{fig-3}. We can then immediately relate the unit vectors in the $1'2'3'$-system to those in the $1''2''3''$-system:
\begin{eqnarray}
{\mathbf u}'_1 &=& \cos \theta {\mathbf u}''_1 + \sin \theta {\mathbf u}''_3   \nonumber \\
{\mathbf u}'_2 &=& {\mathbf u}''_2 \label{eq-ss-8f} \\
{\mathbf u}'_3 &=& - \sin \theta {\mathbf u}''_1 + \cos \theta {\mathbf u}''_3 . \nonumber 
\end {eqnarray}
and
\begin{eqnarray}
{\mathbf u}''_1 &=& \cos \theta {\mathbf u}'_1 - \sin \theta {\mathbf u}'_3   \nonumber \\
{\mathbf u}''_2 &=& {\mathbf u}'_2 \nonumber \\
{\mathbf u}''_3 &=& \sin \theta {\mathbf u}'_1 + \cos \theta {\mathbf u}'_3 . \label{eq-ss-9f} 
\end {eqnarray}

\subsection{\protect\bigskip Current Matrix Elements\label{subsec-currentme}}

The discussions of the on-shell single-nucleon current to follow all come from the study presented in \cite{JandD}. We begin by transforming some of the results presented there into the present notation. In particular, in that cited work the coordinate system was chosen to be referenced to the hadron plane and accordingly what are called $({\mathbf u}_1,{\mathbf u}_2,{\mathbf u}_3)$ there are $({\mathbf u}^{\prime}_1,{\mathbf u}^{\prime}_2,{\mathbf u}^{\prime}_3)$ in the present choice of conventions. For clarity we rewrite the results in \cite{JandD} using the present notation. The familiar single-nucleon on-shell current matrix element is given as
\begin{equation}
J^{\mu }(P^{\prime }\Lambda ^{\prime };P\Lambda )=\bar{u}(P^{\prime
}\Lambda ^{\prime })\left[ F_1\gamma ^{\mu }+\frac{i}{2 m_N} F_2 \sigma^{\mu\nu}Q_\nu\right] u(P\Lambda ),  \label{sn33v}
\end{equation}
where the initial and final spin projections are denoted $\Lambda$ and $\Lambda'$, respectively, with $\Lambda,\Lambda' = \pm 1/2$. As in \cite{JandD} we introduce the following expression in which a current operator enters sandwiched between two spin-1/2 spinors:
\begin{equation}
J^\mu (P\Lambda;P' \Lambda^\prime) \equiv \chi^\dagger_{\Lambda^\prime} {\bar J}^\mu (P;P') \chi_\Lambda . \label{eq-sn-69} 
\end{equation}
As noted in \cite{JandD}, one sees that while $J^\mu$ in Eq.~(\ref{sn33v}) is a 4-vector, ${\bar J}^\mu$ in Eq.~(\ref{eq-sn-69}) is not. 

Next we summarize results found in \cite{JandD} for explicit expressions involving the operator ${\bar J}^\mu$:
\begin{eqnarray}
{\bar J}^\mu &\equiv& f_0 V^\mu \label{eq-sn-80} \\
f_0 &\equiv& \alpha_1 \alpha_2 {\tilde f}_0 \label{eq-sn-81} \\
{\tilde f}_0 &\equiv& \left( \alpha_2^2 +\frac{\tau}{4(1+\tau)} \delta^2 \right)^{-1/2} . \label{eq-sn-82}
\end{eqnarray}

As in \cite{JandD} we employ dimensionless variables:
\begin{eqnarray}
\lambda &\equiv& \omega/2 m_N \nonumber \\
{\bm \kappa} &\equiv& {\mathbf q}/2 m_N \nonumber \\
\tau &\equiv& \frac{|Q^2|}{4 m_N^2}
= \kappa^2 -\lambda^2 \ge 0 \nonumber \\
\rho &\equiv& \frac{|Q^2|}{q^2} = \frac{\tau}{\kappa^2} \qquad 0 \leq \rho \leq 1 \nonumber \\
{\bm \eta} &\equiv& {\mathbf p}/m_N \nonumber \\
\varepsilon &\equiv& E_p/m_N = \sqrt{1+\eta^2} . \label{eq-sn-74} 
\end{eqnarray}
We have that
\begin{eqnarray}
\delta &\equiv& \eta \sin \theta \nonumber \\
\delta^\prime &\equiv& \eta \cos \theta \nonumber \\
\kappa \delta^\prime &=& \lambda \varepsilon - \tau = \lambda {\bar \varepsilon} - \kappa^2 \label{eq-sn-75} \\
{\bar \varepsilon} &\equiv& \varepsilon + \lambda = \frac{1}{m_N} \cdot \frac{1}{2} \left( E'_p + E_p \right) = \frac{\kappa}{\sqrt{\tau}} \sqrt{ 1 + \tau + \delta^2 } , \label{eq-sn-77}
\end{eqnarray}
where $\theta$ is the angle between ${\mathbf q}$ and ${\mathbf p}$ (see above). 
Additionally, using results from \cite{JandD} we introduce the following:
\begin{eqnarray}
\alpha_1 &\equiv& \frac{1}{\mu_1} = \frac{\sqrt{\tau} {\bar \varepsilon}}{\kappa \sqrt{1 + \tau}}  \nonumber \\
&=& \sqrt{1 + \frac{\delta^2}{1+\tau}} \label{eq-sn-78a} \\
\alpha_2 &\equiv& \frac{1}{\mu_2} = \frac{ \sqrt{\tau} (1 + \tau + {\bar \varepsilon}) }{ 2 \kappa \sqrt{1+\tau} }  \nonumber \\
&=& \frac{1}{2} \left( \alpha_1 + \zeta \right) \label{eq-sn-79a} 
\end{eqnarray}
with
\begin{equation}
\zeta \equiv \frac{\sqrt{\tau (1+\tau)}}{\kappa} . \label{eq-nn45a}
\end{equation}
One can also show that
\begin{equation}
{\tilde f}_0^{-2} = \frac{\tau}{\kappa^2} \Lambda \left( \Lambda + \lambda \right) = \frac{\tau}{4\kappa^2} \left[ \left( 1 + {\bar\varepsilon} \right)^2 -\lambda^2 \right] , \label{eq-nn34}
\end{equation}
where
\begin{equation}
\Lambda \equiv \frac{1}{2} \left( 1 + \varepsilon  \right) . \label{eq-nn35}
\end{equation}

Using results from above we may express the kinematic variables in terms of a single set of three independent quantities; specifically, let us write everything in terms of $(\tau, \eta, \theta)$. We have that $\varepsilon = \sqrt{1+\eta^2}$, $\delta = \eta \sin \theta$ and $\delta' = \eta \cos\theta$, and that the dimensionless 3-momentum transfer and energy transfer may be written
\begin{eqnarray}
\kappa &=& \frac{\sqrt{\tau}}{1+ \delta^2} \left[ \sqrt{\tau} \delta' + \varepsilon \sqrt{1+\tau+\delta^2} \right] \label{eq-nnn1a} \\
\lambda &=& \frac{\sqrt{\tau}}{1+ \delta^2} \left[ \sqrt{\tau} \varepsilon  + \delta'\sqrt{1+\tau+\delta^2} \right] . \label{eq-nnn1b} 
\end {eqnarray}
Other useful identities are the following: from Eq.~(\ref{eq-sn-75}) one can show that
\begin{equation}
\kappa \delta^\prime = \kappa \eta \cos\theta = \lambda \varepsilon - \tau = \left( \lambda -\tau \right) + \frac{\lambda \eta^2}{1+\varepsilon } \label{eq-nnn1c}
\end{equation}
and therefore that
\begin{eqnarray}
\lambda -\tau &=& \kappa \delta^\prime - \frac{\lambda \eta^2}{1+\varepsilon } \label{eq-nnn1d} \\
\lambda - \tau\varepsilon &=& \kappa \delta' - \frac{\left( \tau + \lambda \right) \eta^2}{1+\varepsilon } . \label{eq-nnn1e}
\end {eqnarray}

We then turn to focus on the 4-vector $V^\mu = (V^0, {\mathbf v})$ with ${\mathbf v} = V^{1^\prime} {\mathbf u}^{\prime}_1 + V^{2^\prime} {\mathbf u}^{\prime}_2 + V^{3^\prime} {\mathbf u}^{\prime}_3$ and with $V^\mu \equiv \mu_1 \mu_2 {\widetilde V}^\mu$. From \cite{JandD} we have that
\begin{eqnarray}
V^0 &=& \nu_0 +i \nu'_0 \left( {\bm u}'_2 \cdot {\bm \sigma} \right) \nonumber  \\
V^3 &=& \left( \frac{\lambda}{\kappa} \right) V^0 \nonumber  \\
{\bm v}^{\perp} &=& \nu_1 {\bm u}'_1 \nonumber \\
&&-i \left\{ \nu'_2 \left( {\bm u}'_3 \times {\bm \sigma} \right) + \nu''_2 \left( {\bm u}'_3 \cdot {\bm \sigma} \right) {\bm u}'_2 + \left( \nu'_2 - \nu'_1 \right) \left( {\bm u}'_2 \cdot {\bm \sigma} \right) {\bm u}'_1 \right\} , \label{eq-cur-3}
\end{eqnarray}
and accordingly 
\begin{eqnarray}
V^0 &=& \nu_0 +i\nu'_0 \sigma^{2'} \nonumber \\
V^{1^\prime} &=& \nu_1 + i \nu'_1 \sigma^{2'} \nonumber \\
V^{2^\prime} &=&  -i \left[ \nu'_2 \sigma^{1'} + \nu''_2 \sigma^{3'} \right]\nonumber  \\
V^{3^\prime} &=& V^3 = \left( \frac{\lambda}{\kappa} \right) V^0 , \label{eq-sn-86} 
\end{eqnarray}
the last arising from the continuity equation. 
The functions $\nu_0$, {\it etc.} are all real and have been derived previously in \cite{JandD}. For convenience we also choose to remove factors $\mu_1 \mu_2$ by defining ${\nu}_0 \equiv \mu_1 \mu_2 {\tilde \nu}_0 = {\tilde \nu}_0/\alpha_1 \alpha_2$, {\it etc.} where, 
by doing so to define the functions with tildes, one compensates the factors $\alpha_1 \alpha_2$ in Eq.~(\ref{eq-sn-81}). The required quantities are then found to be the following:
\begin{eqnarray}
{\tilde \nu}_0 &=& \frac{\kappa}{\sqrt{\tau}} \left[ \alpha_1 \alpha_2 G_E + \frac{1}{2(1+\tau)}\tau G_M \delta^2 \right] \nonumber \\
{\tilde \nu}'_0 &=& \frac{\kappa}{\sqrt{1+\tau}} \left[ \alpha_2 G_M - \frac{1}{2} \alpha_1 G_E \right] \delta \nonumber \\
{\tilde\nu}_1 &=& \frac{1}{\sqrt{1+\tau}} \left[ \alpha_2 G_E + \frac{1}{2} \alpha_1 \tau G_M \right] \delta\nonumber \\
{\tilde\nu}'_1 &=& \sqrt{\tau} \left[ \alpha_1 \alpha_2 G_M - \frac{1}{2(1+\tau)} G_E \delta^2 \right]\nonumber \\
{\tilde\nu}'_2 &=& \sqrt{\tau} \left[ \alpha_1 \alpha_2 - \frac{1}{2(1+\tau)} \delta^2 \right] G_M \nonumber \\
{\tilde\nu}''_2 &=& \frac{1}{2} \left( \frac{\lambda}{\kappa} \right) \sqrt{\tau} G_M \delta . \label{eq-sn-92} 
\end{eqnarray}

Substituting from above for the coefficients we then obtain the $V^0$ and $V^3$ components of the current,
 \begin{eqnarray}
V^0 &=& \frac{\kappa}{\sqrt{\tau}} \left[ G_E +\frac{1}{2\alpha_1 \alpha_2} \left( 2 \alpha_2 G_M - \alpha_1 G_E \right) i\left( {\bm u}'_2 \cdot {\bm \sigma} \right) \cdot \Xi +\frac{1}{2\alpha_1 \alpha_2} G_M \cdot \Xi^2 \right] \label{eq-cur-1-zz}  \\
V^{3'} &=& V^3 = \left( \frac{\lambda}{\kappa} \right) V^0 \label{eq-cur-2-zz} 
\end{eqnarray}
where 
\begin{equation}
\Xi \equiv \sqrt{\frac{\tau}{1+\tau}} \delta , \label{eq-del}
\end{equation}
and, with ${\bm v}^{\perp} = V^{1'} {\bm u}'_1 + V^{2'} {\bm u}'_2$, the transverse components of the 3-vector current
\begin{eqnarray}
V^{1'} &=& \frac{1}{\sqrt{\tau}} \left[ \tau G_M i \left( {\bm u}'_2 \cdot {\bm \sigma} \right) + \frac{1}{2\alpha_1 \alpha_2} \left( 2\alpha_2 G_E + \alpha_1 \tau G_M \right) \cdot \Xi \right. \nonumber \\
&&\left.  \qquad\qquad\qquad\qquad - \frac{1}{2\alpha_1 \alpha_2} G_E i \left( {\bm u}'_2 \cdot {\bm \sigma} \right) \cdot \Xi^2 \right] \nonumber  \\
V^{2'} &=& -i \sqrt{\tau} G_M \left[ \left( {\bm u}'_1 \cdot {\bm \sigma} \right)  + \frac{1}{2\alpha_1 \alpha_2} \left( \frac{\lambda}{\kappa} \right) \sqrt{ \frac{1+\tau}{\tau}} \left( {\bm u}'_3 \cdot {\bm \sigma} \right) \cdot \Xi \right. \nonumber \\ &&\left. \qquad\qquad\qquad\qquad - \frac{1}{2\tau\alpha_1 \alpha_2} \left( {\bm u}'_1 \cdot {\bm \sigma} \right) \cdot \Xi^2 \right]  . \label{eq-cur-2-qq}
\end{eqnarray}
These results are completely general, namely, {\it no expansions in any of the kinematic variables have been made; they may be used at any energy scale}. As above we may take as independent variables $(\tau,\eta,\theta)$ and via $\delta \equiv \eta \sin\theta$ together with the relationships for $\kappa$ and $\lambda$ in Eqs.~(\ref{eq-nnn1a}) and (\ref{eq-nnn1b}), respectively, determine all of the needed kinematic factors. These expressions may be used in ultra-relativistic situations such as in collider physics where one may have a high-energy electron colliding with a high-energy proton; both particles may have energies and momenta involving relativistic $\gamma$-variables much larger than unity. For instance, if one has a proton of 250 GeV colliding with an energetic electron then $\eta \cong 266$ and therefore $\delta = \eta \sin\theta$, which enters above in the quantity $\Xi$, becomes unity at a scattering angle of about 0.22$^o$. Clearly for scattering at larger angles $\Xi$ cannot be taken to be small and all contributions to $V^\mu$ above must be taken into account. Finally, note that the developments in the present work are completely compatible with those undertaken in \cite{Sofiatti:2011yi} where the focus was entirely placed on collider physics. The present study is focused more on the general $(\tau,\eta,\theta)$-dependencies of the currents and later the single-nucleon response functions.

That said, it is frequently useful to make some reasonable approximations. As mentioned above there are circumstances where we may assume that $\eta$ is small and accordingly immediately drop the contributions proportional to $\Xi^2$ in the above equations. Moreover, for the terms above that are linear in $\eta$, namely, those having explicit factors of $\Xi$ we may evaluate the rest of the expressions at $\eta \rightarrow 0$, which implies that $\alpha_1$ and $\alpha_2$ go to unity and $\lambda / \kappa \rightarrow \sqrt{ \tau / ( 1+ \tau )}$ which leads to the following linear-order results:
\begin{eqnarray}
V^0 &=& \frac{\kappa}{\sqrt{\tau}} \left[ G_E +\left( G_M - \frac{1}{2} G_E \right) i\left( {\bm u}'_2 \cdot {\bm \sigma} \right) \cdot \Xi + {\cal O}(\eta^2) \right] \label{eq-cur-1-ww} \\
V^{3'} &=& V^3 = \left( \frac{\lambda}{\kappa} \right) V^0 \label{eq-cur-2-ww} \\
V^{1'} &=& \frac{1}{\sqrt{\tau}} \left[ \tau G_M i \left( {\bm u}'_2 \cdot {\bm \sigma} \right) + \left(  G_E + \frac{1}{2} \tau G_M \right) \cdot \Xi  + {\cal O}(\eta^2) \right] \label{eq-cur-3-ww} \\
V^{2'} &=& -i \sqrt{\tau} G_M \left[ \left( {\bm u}'_1 \cdot {\bm \sigma} \right)  + \frac{1}{2} \left( {\bm u}'_3 \cdot {\bm \sigma} \right) \cdot \Xi + {\cal O}(\eta^2) \right]  , \label{eq-cur-4-ww}
\end{eqnarray} 
where we have chosen to retain the factor $\kappa / \sqrt{\tau}$ in Eq.~(\ref{eq-cur-1-ww}) although, as stated above, it may also be expanded in $\eta$ (see Eq.~(\ref{eq-nn57})):
\begin{equation}
\frac{\kappa}{\sqrt{\tau}} = \sqrt{1+\tau} \left[ 1 + \sqrt{ \frac{\tau}{1+\tau} } \eta \cos\theta + {\cal O} (\eta^2) \right]. \label{eq-nn57-z}
\end{equation}
See also the discussions to follow in Sec.~\ref{sec-num}. We note that the leading-order contributions (${\cal O}(\eta^0)$) are proportional to $G_E$ for the charge sector and to $G_M$ for the transverse current sector, the latter involving the spin operator via the factors ${\bm u}'_1 \cdot {\bm \sigma}$ and ${\bm u}'_2 \cdot {\bm \sigma}$. That is, the leading-order effects arise via the charge operator and the transverse projections of the spin current operator. The linear contributions (${\cal O}(\eta^1)$) involve both $G_E$ and $G_M$, which is natural, since one expects charge-like effects to arise from a moving magnetic moment, and magnetic-like effects to arise from a moving charge. 

In Eq.~(\ref{eq-cur-1-ww}) we see that these contributions are proportional to $G_M - \frac{1}{2} G_E$ and involve the spin operator via ${\bm u}'_2 \cdot {\bm \sigma}$. In Eq.~(\ref{eq-cur-3-ww}) the linear-order contribution goes as $G_E + \frac{1}{2} \tau G_M$ and is spin-independent --- this is the convection current. Moreover, in Eq.~(\ref{eq-cur-4-ww}) the leading-order contribution is spin-dependent, involving the factor ${\bm u}'_1 \cdot {\bm \sigma}$, together with a first-order contribution that is proportional to $G_M$ and involves the spin operator via the factor ${\bm u}'_3 \cdot {\bm \sigma}$. Indeed, note that a strict expansion in $\eta$ would yield only a magnetization current in leading order and only at ${\cal O}(\eta)$ would one have a convection current, while, of course, most treatments include the latter. However, some treatments of the single-nucleon EM current include {\it only} the convection current at linear order in $\eta$ and do not include the linear-order contributions to the charge operator and to the transverse current contribution in Eq.~(\ref{eq-cur-4-ww}). This is inconsistent and, depending on the specific kinematics these linear-order contributions are not insignificant (see later).

Note that we have not expanded in $\omega/2 m_N \equiv \lambda$, $q/2 m_N \equiv \kappa$ or $|Q^2|/4 m_N^2 \equiv \tau$, but have assumed that these kinematic quantities may take on any values and, in general, should not be assumed to be small and therefore expandable to low orders. All that has been assumed is that an expansion in $p/ m_N \equiv \eta$ may be made, and even this is not necessary if the goal is simply to have results for the on-shell single-nucleon current matrix elements, since above we have the exact results to all orders.

However, there are many circumstances when using these results in nuclear physics applications where various expansions are typically undertaken. We advocate using Eqs.~(\ref{eq-cur-1-ww}--\ref{eq-cur-4-ww}) in general, since there are important occasions where only an expansion in $\eta$ makes sense. Such expansions in $\eta$ do make sense, for instance, when treating quasielastic electron scattering in circumstances where the main contributions arise from nucleons in the nucleus that are relatively slowly-moving, namely, where $p \ll m_N$. We shall discuss this in more detail in Sec.~\ref{sec-num}. However, in many modern applications of interest in quasielastic electron scattering one certainly cannot assume that $\lambda$, $\kappa$ and $\tau$ may be treated as small. Accordingly, the basic ``prescription for nuclear physics'' may be taken to be the following: start with Eqs.~(\ref{eq-cur-1-ww}--\ref{eq-cur-4-ww}) and, where $\eta$ enters via the factors $\delta \equiv \eta \sin\theta = p \sin\theta / m_N$, which are part of the definition of $\Delta$, and replace ${\bm p}$ with $-i {\bm \nabla}$. This prescription is then employed in coordinate space where one has to evaluate matrix elements both of the operator ${\bm \nabla}$ and of the spin operator ${\bm \sigma}$ --- these are all familiar ideas in standard treatments of electromagnetic nuclear physics using (typically) non-relativistic nuclear many-body wave functions.

Let us discuss  in a little more detail several examples of what this approach entails. First, consider $V^0$ in Eq.~(\ref{eq-cur-1-zz}) and to make estimates of how relevant the linear-order contribution is with respect to the leading-order contribution consider the following approximations for the nucleon form factors when the focus is on the {\it isovector} sector. This sector is of special interest since typically it proves to be important to analyze isovector electron scattering before going on to model charge-changing neutrino reactions, which are closely related processes and where the latter are entirely isovector in nature. Take $G_E^v = G_D$ and $G_M^v = \mu^v G_D$ with $\mu^v = \mu_p - \mu_n = 4.70$ for the isovector magnetic moment, where $G_D$ is the dipole form factor often used as a (reasonably good) approximation for the measured form factors 
(see \cite{Donnelly:1978tz,Musolf:1993tb}). Accordingly, the leading-order contribution is proportional to $G_E^v = G_D$, while the next-to-leading-order contribution is proportional to $G_M^v - \frac{1}{2} G_E^v = 4.20 G_D$. If one's interest is in multi-GeV quasielastic electron scattering the dependence on $\tau$ cannot be assumed to be small --- taking $\tau \sim$0.25-0.5 is typical --- and therefore the basic measure is to compare the leading order $\sim$1 with the next-to-leading-order $\sim $4.20$\eta$. As noted above $\eta$ typically lies somewhat below $\eta_F \approx$ 0.25 and thus the linear contribution becomes $\sim$1/4 and one should therefore expect roughly 25\% effects from the contributions that are ${\cal O}(\eta^1)$ versus those of ${\cal O}(\eta^0)$; going the other way, contributions of ${\cal O}(\eta^2)$ may be only of order a few percent. These estimates are consistent with the results for the response functions obtained later in this work.

Another example that occurs in the quasielastic regime is when neutrons are involved. Then one has $G_M^n \approx \mu_n G_D$ along with the Galster parametrization for the neutron's electric form factor, $G_E^n \approx - \mu_n \tau G_D (1 + 5.6 \tau)^{-1}$ (see \cite{Donnelly:1978tz,Musolf:1993tb}). In this case, at moderate values of $\tau$ the leading-order contribution is strongly suppressed with respect to the linear-order contribution, and any modeling of quasielastic electron scattering from neutrons that lacks both terms is liable to be incorrect for the charge sector.

Finally, we briefly note that the above results expanded to linear order in $\eta$ agree with other ``prescriptions for nuclear physics", such as those in \cite{DeForest:1966ycn} (see Appendix D in that work as well as \cite{McVoy}), once the different assumptions made there are taken into account. Specifically, note that in \cite{DeForest:1966ycn}  the direction of the momentum transfer is opposite to the one chosen in this work ($Q^\mu_{DW} = - Q^\mu$) and that the spinor conventions are also different. Moreover, those authors expand in powers of $m_M^{-1}$, retaining contributions only up to ${\cal O}(m_N^{-2})$. They also assume that $\omega \sim m_N^{-1}$, which is valid for elastic scattering from nuclei or excitation of low-lying states in nuclei, and hence $\lambda \sim m_N^{-2}$ rather that being of ${\cal O}(m_N^{-1})$ as is the case for high-energy quasielastic scattering. Our focus in the present work has been on this latter regime and hence we have a contribution that is absent even when expanding in powers of $m_N^{-1}$. Nevertheless, the final expressions are very similar if these differences are taken into account and if one maintains the distinction between 3- and 4-momentum transfers.

\section{\protect\bigskip Single-Nucleon Electron Scattering Response Functions\label{sec-elscatt}}

We next turn to the single-nucleon (elastic) response functions. We again build on the study undertaken in \cite{JandD}; there only the unpolarized responses were considered and thus the present work constitutes an extension to include the situation where the target nucleon is polarized. We begin with a few developments concerning general EM hadronic tensors before specializing to the on-shell single-nucleon case.

\subsection{\protect\bigskip General Electromagnetic Hadronic Tensors\label{subsec-tensors}}

In this section we discuss the general form of the hadronic tensor $W^{\mu\nu}$ used in studies of electromagnetic interactions with nucleons and nuclei, again drawing on the presentations in \cite{Donnelly:2023rej,arxivlong}. This satisfies the continuity equation:
\begin{equation}
Q_\mu W^{\mu\nu} = Q_\nu W^{\mu\nu} = 0 . \label{eq-rot-15}
\end{equation}
With our conventions where ${\mathbf q}$ lies in the 3-direction,
\begin{equation}
Q^\mu = (\omega, 0,0,q) , \label{eq-rot-16}
\end{equation}
and we can then eliminate any tensors with longitudinal ($\mu = 3$) components in terms of components having $\mu = 0$:
\begin{eqnarray}
W^{3\mu} &=& \nu^{\prime} W^{0\mu} \nonumber \\
W^{\mu 3} &=& \nu^{\prime} W^{\mu 0} , \label{eq-rot-18}
\end{eqnarray}
where as usual $\nu^{\prime} \equiv \omega / q$. Thus the only cases we need to consider when rotating are $W^{01}$, $W^{10}$, $W^{02}$, $W^{20}$,  $W^{11}$, $W^{22}$, $W^{12}$ and $W^{21}$. 

Now, moving to the EM response functions employed in other work (see \cite{Donnelly:2023rej,arxivlong,Donnelly:1985ry,Raskin:1988kc}), we have the following nine familiar (real) response functions:
\begin{eqnarray}
W^L &=& W^{00} \nonumber \\
W^T &=& W^{22} + W^{11} \nonumber \\
W^{TT} &=& W^{22} - W^{11} \nonumber \\
W^{TL} &=& 2\sqrt{2} W^{01}_s = 2\sqrt{2} {\rm Re}W^{01}  \nonumber \\
W^{T'} &=& 2iW^{12}_a = -2{\rm Im}W^{12} \nonumber \\
W^{TL'} &=& 2\sqrt{2}i W^{02}_a = -2\sqrt{2} {\rm Im}W^{02} \nonumber \\
W^{\underline{TT}} &=& 2 W^{12}_s = 2{\rm Re}W^{12} \nonumber \\
W^{\underline{TL}} &=& 2\sqrt{2} W^{02}_s = 2\sqrt{2} {\rm Re}W^{02} \nonumber \\
W^{\underline{TL}^\prime} &=& -2\sqrt{2}i W^{01}_a = 2\sqrt{2} {\rm Im}W^{01} . \label{eq-rot-50} 
\end{eqnarray}
Here the subscripts ``s'' and ``a'' are used to indicate the parts of the hadronic tensor that are symmetric and anti-symmetric under the interchange $ \mu \leftrightarrow \nu$, respectively. 
We may also introduce similar response functions in the rotated coordinate system, denoted by tildes:
\begin{eqnarray}
{\widetilde W}^L &=& W^{00} \nonumber \\
{\widetilde W}^T &=& W^{{2^\prime}{2^\prime}} + W^{{1^\prime}{1^\prime}} \nonumber \\
{\widetilde W}^{TT} &=& W^{{2^\prime}{2^\prime}} - W^{{1^\prime}{1^\prime}} \nonumber \\
{\widetilde W}^{TL} &=& 2\sqrt{2} W^{0{1^\prime}}_s = 2\sqrt{2} {\rm Re}W^{0{1^\prime}} \nonumber \\
{\widetilde W}^{T'} &=& 2iW^{{1^\prime}{2^\prime}}_a = -2 {\rm Im}W^{{1^\prime}{2^\prime}} \nonumber \\
{\widetilde W}^{TL'} &=& 2\sqrt{2}i W^{0{2^\prime}}_a = -2\sqrt{2} {\rm Im}W^{0{2^\prime}} \nonumber \\
{\widetilde W}^{\underline{TT}} &=& 2 W^{{1^\prime}{2^\prime}}_s = 2 {\rm Re}W^{{1^\prime}{2^\prime}} \nonumber \\
{\widetilde W}^{\underline{TL}} &=& 2\sqrt{2} W^{0{2^\prime}}_s = 2\sqrt{2} {\rm Re}W^{0{2^\prime}} \nonumber \\
{\widetilde W}^{\underline{TL}^\prime} &=& -2\sqrt{2}i W^{0{1^\prime}}_a = 2\sqrt{2} {\rm Im}W^{0{1^\prime}}. \label{eq-rot-59} 
\end{eqnarray}

Upon employing explicit identities relating the various coordinate systems discussed above, the following relationships between the EM responses in the $1 2 3$-system and those in the $1'2'3'$-system emerge immediately:
\begin{eqnarray}
W^L &=& {\widetilde W}^L \nonumber \\
W^T &=& {\widetilde W}^T \nonumber \\
W^{TT} &=& {\widetilde W}^{TT} \cos 2\phi + {\widetilde W}^{\underline{TT}} \sin 2\phi \nonumber \\
W^{TL} &=& {\widetilde W}^{TL} \cos \phi - {\widetilde W}^{\underline{TL}} \sin \phi \nonumber \\
W^{T'} &=& {\widetilde W}^{T'} \nonumber \\
W^{TL'} &=& {\widetilde W}^{TL'} \cos \phi - {\widetilde W}^{\underline{TL}^\prime} \sin \phi \nonumber \\
W^{\underline{TT}} &=& {\widetilde W}^{\underline{TT}} \cos 2\phi  - {\widetilde W}^{TT} \sin 2\phi \nonumber \\
W^{\underline{TL}} &=& {\widetilde W}^{\underline{TL}} \cos \phi  + {\widetilde W}^{TL} \sin \phi\nonumber \\
W^{\underline{TL}^\prime} &=& {\widetilde W}^{\underline{TL}^\prime} \cos \phi +  {\widetilde W}^{TL'} \sin \phi . \label{eq-rot-68} 
\end{eqnarray}
Of course one may derive the inverse relationships; however, for the present purposes the above equations are adequate.

\subsection{\protect\bigskip On-shell Single-nucleon Response \label{subsec-single}}

The discussions in this section again build on the study presented in \cite{JandD}, here with extensions to include single-nucleon responses where the target nucleon is polarized --- the previous study was limited to the unpolarized sector. For the response functions we require the bilinear products of the current matrix elements
\begin{eqnarray}
W^{\mu \nu}_{\Lambda' \Lambda} &\equiv& J^\mu (P\Lambda;P' \Lambda^\prime)^* J^\nu (P\Lambda;P' \Lambda^\prime) \nonumber \\
&=& \left( \chi^\dagger_{\Lambda^\prime} {\bar J}^{\mu} (P;P') \chi_\Lambda \right)^* \left( \chi^\dagger_{\Lambda^\prime} {\bar J}^\nu (P;P') \chi_\Lambda \right) \nonumber\\
&=&  \chi^\dagger_\Lambda {\bar J}^{\mu} (P;P')^\dagger \chi_{\Lambda^\prime} \chi^\dagger_{\Lambda^\prime}  {\bar J}^\nu (P;P') \chi_\Lambda . \label{eq-aa3}
\end{eqnarray}
Since in the present work only the initial incoming nucleon is assumed to be polarized we must sum over final spin states and then, using closure, we obtain the following:
\begin{eqnarray}
W^{\mu \nu}_{\Lambda} &\equiv& \sum_{\Lambda^\prime} W^{\mu \nu}_{\Lambda' \Lambda}  \nonumber  \\
&=& \chi^\dagger_\Lambda {\bar J}^{\mu} (P;P')^\dagger {\bar J}^\nu (P;P') \chi_\Lambda . \label{eq-aaa5}
\end{eqnarray}
Using the results in Eq.~(\ref{eq-aaa5}) we then have
\begin{equation}
W^{\mu \nu}_{\Lambda} =  {\tilde f}^2_0 \chi^\dagger_\Lambda {\widetilde V}^{\mu \dagger} {\widetilde V}^\nu \chi_\Lambda . \label{eq-aa6}
\end{equation}
The various required combinations of ${V^\mu}^\dagger V^\nu$ are the following:
\begin{eqnarray}
{\widetilde V}^{0 \dagger} {\widetilde V}^0 &=& {\tilde \nu}^2_0 + {\tilde \nu}^{\prime 2}_0 \equiv Z_{00} \label{eq-aa7} \\
{\widetilde V}^{1' \dagger} {\widetilde V}^{1'} &=& {\tilde \nu}^2_1 + {\tilde \nu}^{\prime 2}_1 \equiv Z_{1'1'} \label{eq-aa8} \\
{\widetilde V}^{2' \dagger} {\widetilde V}^{2'} &=& {\tilde \nu}^{\prime 2}_2 + {\tilde \nu}^{\prime\prime 2}_2  + {\tilde\nu}'_2 {\tilde\nu}''_2 \left[ \sigma^{1'\dagger} \sigma^{3'} + \left( \sigma^{1'\dagger} \sigma^{3'} \right)^\dagger  \right] \nonumber \\
&=& {\tilde \nu}^{\prime 2}_2 + {\tilde \nu}^{\prime\prime 2}_2 \equiv Z_{2'2'} \label{eq-aa10} \\
{\widetilde V}^{0 \dagger} {\widetilde V}^{1'} &=& \left[ {\tilde \nu}_0 {\tilde \nu}_1 + {\tilde \nu}'_0 {\tilde \nu}'_1  \right] +i\left[ {\tilde \nu}_0 {\tilde \nu}'_1 - {\tilde \nu}'_0 {\tilde \nu}_1  \right] \sigma^{2'} \nonumber \\ &=& Z_{01'} + iZ'_{01'} \sigma^{2'}  \label{eq-aa9} \\
{\widetilde V}^{0 \dagger} {\widetilde V}^{2'} &=& -i \left[ {\tilde \nu}_0 -i{\tilde \nu}'_0 \sigma^{2'} \right] \left[ {\tilde\nu}'_2 \sigma^{1'} + {\tilde\nu}''_2 \sigma^{3'}  \right] \nonumber \\
&=& -i \left\{ \left[ {\tilde \nu}_0 {\tilde\nu}'_2 + {\tilde \nu}'_0 {\tilde\nu}''_2 \right] \sigma^{1'} + \left[ {\tilde \nu}_0 {\tilde\nu}''_2 - {\tilde \nu}'_0 {\tilde\nu}'_2 \right] \sigma^{3'}  \right\} \nonumber \\
&\equiv& -i \left\{ Z_{02'}  \sigma^{1'} + Z'_{02'}  \sigma^{3'}  \right\}  \label{eq-aa11}  \\ 
{\widetilde V}^{1' \dagger} {\widetilde V}^{2'} &=& -i \left[ {\tilde \nu}_{1'} -i{\tilde \nu}'_{1'} \sigma^{2'} \right] \left[ {\tilde\nu}'_2 \sigma^{1'} + {\tilde\nu}''_2 \sigma^{3'}  \right] \nonumber \\
&=& -i \left\{ \left[ {\tilde \nu}_1 {\tilde\nu}'_2 + {\tilde \nu}'_1 {\tilde\nu}''_2 \right] \sigma^{1'} + \left[ {\tilde \nu}_1 {\tilde\nu}''_2 - {\tilde \nu}'_1 {\tilde\nu}'_2 \right] \sigma^{3'}  \right\} \nonumber \\
&\equiv& -i \left\{ Z_{1'2'}  \sigma^{1'} + Z'_{1'2'}  \sigma^{3'}  \right\}  , \label{eq-aa12} 
\end{eqnarray}
where the fact that the Pauli matrices are hermitean has been used, as well as the fact that
\begin{equation}
\sigma^k \sigma^\ell = \delta^{k\ell} +i \epsilon^{k\ell m} \sigma^m , \label{eq-aa13} 
\end{equation}
and where we have defined the following
\begin{eqnarray}
Z_{00} &\equiv& {\tilde \nu}^2_0 + {\tilde \nu}^{\prime 2}_0 \nonumber \\ 
Z_{1'1'} &\equiv& {\tilde \nu}^2_1 + {\tilde \nu}^{\prime 2}_1 \nonumber \\ 
Z_{2'2'} &\equiv& {\tilde \nu}^{\prime 2}_2 + {\tilde \nu}^{\prime\prime 2}_2 \nonumber \\ 
Z_{01'} &\equiv& {\tilde \nu}_0 {\tilde \nu}_1 + {\tilde \nu}'_0 {\tilde \nu}'_1 \nonumber \\ 
Z'_{01'} &\equiv& {\tilde \nu}_0 {\tilde \nu}'_1 - {\tilde \nu}'_0 {\tilde \nu}_1 \nonumber \\ 
Z_{02'} &\equiv& {\tilde \nu}_0 {\tilde\nu}'_2 + {\tilde \nu}'_0 {\tilde\nu}''_2 \nonumber \\ 
Z'_{02'} &\equiv& {\tilde \nu}_0 {\tilde\nu}''_2 - {\tilde \nu}'_0 {\tilde\nu}'_2 \nonumber \\ 
Z_{1'2'} &\equiv& {\tilde \nu}_1 {\tilde\nu}'_2 + {\tilde \nu}'_1 {\tilde\nu}''_2 \nonumber \\ 
Z'_{1'2'} &\equiv& {\tilde \nu}_1 {\tilde\nu}''_2 - {\tilde \nu}'_1 {\tilde\nu}'_2 . \label{eq-aa17}
\end{eqnarray}
Note that the results in Eqs.~(\ref{eq-aa7}--\ref{eq-aa9}) are purely real and therefore contribute only in the (symmetric) unpolarized response sector, while those in Eqs.~(\ref{eq-aa11}--\ref{eq-aa12}) are purely imaginary and therefore contribute only in the (anti-symmetric) unpolarized response sector. Only TRE responses are nonzero. Also note that for the polarized results only those involving the Pauli matrices $\sigma^{1'}$ and $\sigma^{3'}$ occur with none involving $\sigma^{2'}$; the last involving polarizations normal to the plane defined by {\bf q} and {\bf p} are parity violating. Referring to Eqs.~(\ref{eq-rot-59}), these specific results show that only the unpolarized  responses ${\widetilde W}^L$, ${\widetilde W}^T$, ${\widetilde W}^{TT}$ and ${\widetilde W}^{TL}$ are nonzero, whereas ${\widetilde W}^{\underline{TT}}$ and ${\widetilde W}^{\underline{TL}}$ are both zero, and that only the polarized responses ${\widetilde W}^{T'}$ and ${\widetilde W}^{TL'}$ are nonzero, whereas ${\widetilde W}^{\underline{TL'}}$ is zero. The relationships in Eqs.~(\ref{eq-rot-68}) where responses in the $123$-system are related to those in the $1'2'3'$-system then simply reduce to the following:
\begin{eqnarray}
W^L &=& {\widetilde W}^L \nonumber \\ 
W^T &=& {\widetilde W}^T \nonumber \\ 
W^{TT} &=& {\widetilde W}^{TT} \cos 2\phi \nonumber \\ 
W^{TL} &=& {\widetilde W}^{TL} \cos \phi \nonumber \\ 
W^{T'} &=& {\widetilde W}^{T'} \nonumber \\ 
W^{TL'} &=& {\widetilde W}^{TL'} \cos \phi \nonumber \\ 
W^{\underline{TT}} &=&  - {\widetilde W}^{TT} \sin 2\phi \nonumber \\ 
W^{\underline{TL}} &=& {\widetilde W}^{TL} \sin \phi   \nonumber \\ 
W^{\underline{TL}^\prime} &=& {\widetilde W}^{TL'} \sin \phi . \label{eq-rot-68t} 
\end{eqnarray}

To go from these response functions to the relevant electron scattering cross sections one may use the formalism presented in \cite{Donnelly:2023rej,arxivlong} --- all conventions used there are exactly those employed in the present work. From that work one has that the inclusive cross section for scattering of electrons from spin-1/2 targets including electron and target polarizations may be written
\begin{equation}
\frac{d^{2}\sigma }{d\Omega dk^{\prime }}=\sigma _{\mathrm{Mott}}f\frac{m_N}{%
E_{p}}\mathcal{R}^{incl},  \label{eq-long-1}
\end{equation}%
where%
\begin{equation}
\mathcal{R}^{incl}=R_{1}^{incl} + h R_{2}^{incl} + h^* R_{3}^{incl} + hh^* R_{4}^{incl}  \label{eq-long-2}
\end{equation}%
and where $h=\pm1$ and $h^* =\pm1$ carry the polarizations of the electron and nucleon, respectively. The Mott cross section $\sigma _{\mathrm{Mott}}$ and the factor $f$ are discussed in \cite{Donnelly:2023rej,arxivlong}. In the present situation the four sectors (which may be separated by using the polarizations) are found to be%
\begin{eqnarray}
R_{1}^{incl} &=& v_{L}W^{L} + v_{T}W^{T} + v_{TT}W^{TT} + v_{TL}W^{TL} \nonumber \\ 
R_{4}^{incl} &=& v_{T'}W^{T'} + v_{TL'}W^{TL'} \nonumber \\ 
R_{2}^{incl} &=& R_{3}^{incl} = 0 . \label{eq-long-5} 
\end{eqnarray}%

\subsection{\protect\bigskip Unpolarized Single-nucleon Response \label{subsec-unpol}}

For the unpolarized responses (all symmetric in $\mu \leftrightarrow \nu$) in the $1'2'3'$-system the spinor matrix elements are all unity since no Pauli matrices are present and accordingly one may evaluate the spin-averaged responses,
\begin{equation}
W^{\mu\nu}_{unpol} \equiv \frac{1}{2} \left[ W^{\mu \nu}_{+1/2} + W^{\mu \nu}_{-1/2}  \right] ,\label{eq-bb1}
\end{equation}
where immediately we find the following nonzero results
\begin{eqnarray}
W^{00}_{unpol} &=& {\tilde f}_0^2 Z_{00} \nonumber \\ 
W^{1'1'}_{unpol} &=& {\tilde f}_0^2 Z_{1'1'} \nonumber \\ 
W^{2'2'}_{unpol} &=& {\tilde f}_0^2 Z_{2'2'} \nonumber \\ 
W^{01'}_{unpol} &=& W^{1'0}_{unpol} = W^{01' *}_{unpol} = {\rm Re}W^{01'}_{unpol} = {\tilde f}_0^2 Z_{01'} ,\label{eq-sn-132} 
\end{eqnarray}
with no contribution coming from the term proportional to $Z'_{01'}$. These results have already been presented in \cite{JandD}; specifically one has
\begin{eqnarray}
W^{00}_{unpol} &=& {\tilde f}_0^2 Z_{00} = \frac{\kappa^2}{\tau} \left[ G_E^2 + \delta^2 W_2 \right] \nonumber \\ 
W^{1'1'}_{unpol} &=& {\tilde f}_0^2 Z_{1'1'} = W_1 + \delta^2 W_2 \nonumber \\ 
W^{2'2'}_{unpol} &=& {\tilde f}_0^2 Z_{2'2'} = W_1 \nonumber \\ 
W^{01'}_{unpol} &=& {\tilde f}_0^2 Z_{01'} = {\bar \varepsilon} \delta W_2 , \label{eq-sn-132x} 
\end{eqnarray}
where as usual $W_1 = \tau G_M^2$ and $W_2 = \left[ G_E^2 + \tau G_M^2 \right] / (1+\tau)$. Then, using the results in Eqs.~(\ref{eq-rot-59}) in the $1'2'3'$-system one immediately has that
\begin{eqnarray}
{\widetilde W}^L &=& W_{unpol}^{00} = \frac{\kappa^2}{\tau} \left[ G_E^2 + \delta^2 W_2 \right]  \nonumber \\ 
{\widetilde W}^T &=& W_{unpol}^{{2^\prime}{2^\prime}} + W_{unpol}^{{1^\prime}{1^\prime}} = 2W_1 + \delta^2 W_2  \nonumber \\ 
{\widetilde W}^{TT} &=& W_{unpol}^{{2^\prime}{2^\prime}} - W_{unpol}^{{1^\prime}{1^\prime}} = -\delta^2 W_2 \nonumber \\ 
{\widetilde W}^{TL} &=& 2\sqrt{2} W_{unpol}^{0{1^\prime}} = 2\sqrt{2} {\bar \varepsilon} \delta W_2 , \label{eq-rot-54s} 
\end{eqnarray}
together with the facts that 
\begin{equation}
{\widetilde W}^{\underline{TT}} ={\widetilde W}^{\underline{TT}} = 0 , \label{eq-eq-rots1}
\end{equation}
since in  Eqs.~(\ref{eq-rot-59})  where these two responses are defined require the real parts of specific components of the hadronic tensor and yet those components are manifestly imaginary (see Eqs.~(\ref{eq-aa11}--\ref{eq-aa12})). Finally, via Eqs.~(\ref{eq-rot-68}) that one has the required results in the $123$-system, namely, the results already found in \cite{JandD}:

\begin{empheq}[box=\fbox]{align}
W^L &= \frac{\kappa^2}{\tau} \left[ G_E^2 + \delta^2 W_2 \right]  \label{eq-rot-60s} \\
W^T &= 2W_1 + \delta^2 W_2  \label{eq-rot-61s} \\
W^{TT} &=  -\delta^2 W_2 \cos 2\phi  \label{eq-rot-62s} \\
W^{TL} &= 2\sqrt{2} {\bar \varepsilon} \delta W_2  \cos \phi . \label{eq-rot-63s}
\end{empheq}

Note that there are symmetries under the transformation $\theta \rightarrow \pi - \theta$ that emerge immediately: since $\delta = \eta \sin\theta$, under this transformation $\delta \rightarrow \delta$ and therefore $W^{T,TT,TL}$ all do not change, whereas $W^L$ does change because of the transformation of $\kappa$. By examining Eq. ~(\ref{eq-nnn1a}) and noting that $\delta' = \eta \cos\theta \rightarrow - \delta'$ under this transformation one sees that the values of $\kappa$ for $\theta$ and for $\pi - \theta$ are different.

We also note in passing that Eqs.~(\ref{eq-rot-60s}) and (\ref{eq-rot-61s}) were long ago used without approximation in formulating the relativistic Fermi gas model for the longitudinal and transverse responses; see \cite{Alberico:1988bv}.

\subsection{\protect\bigskip Polarized Single-nucleon Response \label{subsec-pol}}

For the case where the initial nucleon is polarized one can employ the following
\begin{equation}
W^{\mu\nu}_{pol} \equiv \frac{1}{2} \left[ W^{\mu \nu}_{+1/2} - W^{\mu \nu}_{-1/2}  \right] , \label{eq-bb2}
\end{equation}
and furthermore, the required response functions expressed in both the $123$- and $1'2'3'$-systems are given in Eqs.~(\ref{eq-rot-50}--\ref{eq-rot-68}). The polarized responses are all anti-symmetric in $\mu \leftrightarrow \nu$; see Eqs.~(\ref{eq-rot-68}), and  
we note that only the $T'$ and $TL'$ responses enter for the polarized responses, the former involving the $1'2'$ and $02'$ Lorentz components of the polarized tensor above. 
In Sec.~\ref{sec-curr} we noted that when considering the spin of the target, it is useful to employ a coordinate system determined by the direction of the incoming nucleon, {\it i.e.,} using the 3-momentum ${\mathbf p}$; we do so now by relating the unit vectors in the $1'2'3'$-system to those in the $1''2''3''$-system, namely where ${\mathbf u}_p \equiv {\mathbf u}''_3$ (see Eqs.~(\ref{eq-ss-9f})). We have already seen that only linear combinations of $\sigma^{1'}$ and $\sigma^{3'}$ enter in the polarized sector and accordingly it is advantageous to rotate these combinations as above, yielding for the components of the Pauli matrices the following relationships for the 3-vector components:
\begin{eqnarray}
\sigma^{1'} &=& \cos \theta \sigma^{1''} + \sin \theta \sigma^{3''}  \nonumber \\ 
\sigma^{2'} &=& \sigma^{2''} \nonumber \\ 
\sigma^{3'} &=& - \sin \theta \sigma^{1''} + \cos \theta \sigma^{3''} . \label{eq-ss-9g} 
\end {eqnarray}
Accordingly, the required bilinear products of the form ${\widetilde V}^{\mu \dagger} {\widetilde V}^\nu$ given in Eqs.~(\ref{eq-aa11}) and (\ref{eq-aa12}) involve the functions
\begin{eqnarray}
Z^{3''}_{02'} &\equiv& Z_{02'} \sin\theta + Z'_{02'} \cos\theta \nonumber \\
Z^{1''}_{02'} &\equiv& Z_{02'} \cos\theta - Z'_{02'} \sin\theta \nonumber \\
Z^{3''}_{1'2'} &\equiv& Z_{1'2'} \sin\theta + Z'_{1'2'} \cos\theta \nonumber \\
Z^{1''}_{1'2'} &\equiv& Z_{1'2'} \cos\theta - Z'_{1'2'} \sin\theta . \label{eq-Z1} 
\end{eqnarray}
Then we require spinor matrix elements of the form
$\chi^\dagger_\Lambda {\widetilde V}^{\mu \dagger} {\widetilde V}^\nu \chi_\Lambda$, where $\mu\nu = 02'$ and $1'2'$. Longitudinal polarization (L) is when the spin axis of quantization is defined to be in the $3''$ direction and sideways polarization (S) when it is defined to be in the $1''$ direction. In the former case one has $\chi^\dagger_\Lambda \sigma^{3''} \chi_\Lambda = 2\Lambda$ and $\chi^\dagger_\Lambda \sigma^{1''} \chi_\Lambda = 0$, while in the latter case one has $\chi^\dagger_\Lambda \sigma^{3''} \chi_\Lambda = 0$ and $\chi^\dagger_\Lambda \sigma^{1''} \chi_\Lambda = 2\Lambda$. We then have for the two types of polarization the following:
\begin{eqnarray}
{\widetilde W}^{TL'}_L &=& 2\sqrt{2} {\tilde f}_0^2 Z^{3''}_{02'} \nonumber \\ 
{\widetilde W}^{TL'}_S &=& 2\sqrt{2} {\tilde f}_0^2 Z^{1''}_{02'} \nonumber \\ 
{\widetilde W}^{T'}_L &=& 2 {\tilde f}_0^2 Z^{3''}_{1'2'}  \nonumber \\ 
{\widetilde W}^{T'}_S &=& 2 {\tilde f}_0^2 Z^{1''}_{1'2'}  \label{eq-polZ4} 
\end{eqnarray}
and therefore the required polarized response functions
\begin{eqnarray}
W^{TL'}_L  &=& 2\sqrt{2} \frac{h^*\kappa}{\zeta} G_M \left[ \alpha_1 G_E - \sqrt{\frac{\tau}{1 +\tau}} G_M \frac{\eta}{\varepsilon} \left( \cos\theta -\frac{\tau}{\kappa} \eta \right) \right]  \sin\theta \cos\phi \nonumber \\ 
&=& 2\sqrt{2} \frac{h^*\kappa}{\zeta} G_M \left[ \alpha_1 G_E - \sqrt{\frac{\tau}{1 +\tau}} G_M \frac{1}{\kappa} \left( \lambda - \tau\varepsilon \right) \right] \sin\theta \cos\phi \nonumber \\ 
&=& 2\sqrt{2} h^*\kappa G_M \frac{1}{1+\tau} \left[ {\bar\varepsilon} G_E - \left( \lambda - \tau\varepsilon\right) G_M \right] \sin\theta \cos\phi \nonumber \\ 
W^{TL'}_S   &=& 2\sqrt{2}  \frac{h^*\kappa}{\zeta} G_M \left[ \frac{\alpha_1}{\varepsilon} \left( \cos\theta -\frac{\tau}{\kappa} \eta \right)G_E + \sqrt{\frac{\tau}{1 +\tau}} G_M \eta \sin^2 \theta \right]  \cos\phi \nonumber \\ 
&=& 2\sqrt{2}  h^*\kappa G_M \frac{1}{(1+\tau) \varepsilon} \left[ {\bar\varepsilon}  \left( \cos\theta -\frac{\tau}{\kappa} \eta \right) G_E + \kappa G_M \varepsilon \eta \sin^2 \theta \right]  \cos\phi \nonumber \\ 
W^{T'}_L   &=&  -2\frac{h^*}{\zeta} G_M \left[ \frac{\alpha_1}{\varepsilon} \left( \cos\theta -\frac{\tau}{\kappa} \eta \right) \tau G_M - \sqrt{\frac{\tau}{1 +\tau}} G_E \eta \sin^2 \theta \right] \nonumber \\ 
&=& -2 G_M \frac{h^*}{(1+\tau) \varepsilon} \left[ {\bar\varepsilon}  \left( \cos\theta -\frac{\tau}{\kappa} \eta \right) \tau G_M - \kappa G_E \varepsilon \eta \sin^2 \theta \right] \nonumber \\ 
W^{T'}_S   &=& 2\frac{h^*}{\zeta} G_M \sin\theta \left[ \alpha_1 \tau G_M + \sqrt{\frac{\tau}{1 +\tau}} G_E \frac{\eta}{\varepsilon} \left( \cos\theta -\frac{\tau}{\kappa} \eta \right) \right] \nonumber \\ 
&=& 2\frac{h^*}{\zeta} G_M \sin\theta \left[ \alpha_1 \tau G_M + \sqrt{\frac{\tau}{1 +\tau}} G_E \frac{1}{\kappa} \left( \lambda - \tau\varepsilon \right) \right] \nonumber \\ 
&=& 2h^* G_M \sin\theta \frac{1}{1+\tau} \left[ {\bar\varepsilon} \tau G_M + \left( \lambda - \tau\varepsilon \right) G_E \right] , \label{eq-nn53b} 
\end {eqnarray}
where we have incorporated the ``spin flipper'' $h^*=\pm 1$ as in \cite{Donnelly:2023rej,arxivlong}. We may also employ the usual relationships between the Sachs form factors and the Pauli/Dirac form factors,
\begin{eqnarray}
G_E &=& F_1-\tau F_2 \nonumber \\ 
G_M &=& F_1+ F_2 , \label{eq-nn2} 
\end {eqnarray}
together with the inverses
\begin{eqnarray}
F_1 &=& \frac{1}{1+\tau} \left( G_E + \tau G_M \right) \nonumber \\ 
F_2 &=& \frac{1}{1+\tau} \left( G_M - G_E \right) , \label{eq-nn4} 
\end {eqnarray}
to write the following very compact forms for the polarized single-nucleon responses:
\begin{empheq}[box=\fbox]{align}
W^{TL'}_L  &= 2\sqrt{2} h^*\kappa G_M \left[ \varepsilon G_E - \left( \lambda - \tau\varepsilon \right) F_2 \right] \sin\theta \cos\phi \label{eq-nn50cE} \\
W^{TL'}_S   &= 2\sqrt{2}  h^*\kappa G_M \left[ \cos\theta G_E + \kappa \eta \sin^2 \theta  F_2 \right] \cos\phi \label{eq-nn51bE} \\
W^{T'}_L   &= -2 h^* G_M \left[ \cos\theta \tau G_M - \kappa \eta \sin^2 \theta  F_1 \right] \label{eq-nn52bE} \\
W^{T'}_S   &= 2h^* G_M \sin\theta \left[ \varepsilon \tau G_M + \left( \lambda - \tau\varepsilon \right) F_1 \right] . \label{eq-nn53cE} 
\end {empheq}
These are new results and are completely general, {\it i.e.,} they are valid for all kinematic conditions and no assumptions have been made (see later) on any of the kinematic variables being small (or for that matter, large).

Note that for $\theta = 0,\pi$ one has the following:
\begin{eqnarray}
\left[ W^{TL'}_L \right]_{0,\pi} &=& \left[ W^{T'}_S \right]_{0,\pi} = 0 \nonumber \\ 
\left[ W^{TL'}_S \right]_{0,\pi} &=& \pm 2\sqrt{2}  h^*\kappa G_E G_M \cos\phi \nonumber \\ 
\left[ W^{T'}_S \right]_{0,\pi} &=& \mp 2 h^* \tau G^2_M , \label{eq-zero4} 
\end{eqnarray}
with the signs being determined by the factor $\cos\theta$ in Eqs.~(\ref{eq-nn51bE}) and (\ref{eq-nn52bE}).
The results in Eqs.~(\ref{eq-nn50cE}--\ref{eq-nn53cE}) may then be combined by employing the polarizations defined by
\begin{eqnarray}
{\cal P}_{1''} &\equiv& \sin\theta^{*\prime \prime} \cos\phi^{*\prime \prime} \nonumber \\ 
{\cal P}_{2''} &\equiv& \sin\theta^{*\prime \prime} \sin\phi^{*\prime \prime} \nonumber \\ 
{\cal P}_{3''} &\equiv& \cos\theta^{*\prime \prime}  \label{eq-sp-3} 
\end{eqnarray}
involving the polarization angles in the $1''2''3''$-system. This leads to the following:
\begin{eqnarray}
W^{TL'}  &=& W^{TL'}_L {\cal P}_{3''} + W^{TL'}_S {\cal P}_{1''} \nonumber \\ 
W^{T'}  &=& W^{T'}_L {\cal P}_{3''} + W^{T'}_S {\cal P}_{1''} .  \label{eq-sp-5} 
\end{eqnarray}
Moreover, it is then possible to re-express these results in terms of polarizations given with respect to the other choices of coordinate system made in this work. Specifically, defining those polarizations by 
\begin{eqnarray}
{\cal P}_{1'} &\equiv& \sin\theta^{*\prime} \cos\phi^{*\prime} \nonumber \\ 
{\cal P}_{2'} &\equiv& \sin\theta^{*\prime} \sin\phi^{*\prime} \nonumber \\ 
{\cal P}_{3'} &\equiv& \cos\theta^{*\prime}  \label{eq-sp-3a} 
\end{eqnarray}
for the $1'2'3'$-system, and
\begin{eqnarray}
{\cal P}_{1} &\equiv& \sin\theta^{*} \cos\phi^{*} \nonumber \\ 
{\cal P}_{2} &\equiv& \sin\theta^{*} \sin\phi^{*} \nonumber \\ 
{\cal P}_{3} &\equiv& \cos\theta^{*}  \label{eq-sp-3b} 
\end{eqnarray}
for the $123$-system, one can use the following identities to relate the results in the three frames:
\begin{eqnarray}
{\cal P}_{1''} &=& \cos\theta {\cal P}_{1'} - \sin\theta {\cal P}_{3'} \nonumber \\ 
&=& \cos\theta \left[ \cos\phi {\cal P}_{1} + \sin\phi {\cal P}_{2} \right] - \sin\theta {\cal P}_{3}  \nonumber \\ 
{\cal P}_{2''} &=& {\cal P}_{2'} \nonumber \\ 
&=& -\sin\phi {\cal P}_{1} +\cos\phi {\cal P}_{2} \nonumber \\ 
{\cal P}_{3''} &=& \sin\theta {\cal P}_{1'} + \cos\theta {\cal P}_{3'}  \nonumber \\ 
&=& \sin\theta \left[ \cos\phi {\cal P}_{1} + \sin\phi {\cal P}_{2} \right] + \cos\theta {\cal P}_{3}  \label{eq-sp-3cx}  
\end{eqnarray}
together with their inverses.

We end this section by specializing to collinear kinematics. As is well-known in studying elastic electron scattering including from nucleons as targets, it proves useful to cast the results in the so-called Breit frame (``B''). In this frame one has the target 3-momentum {\bf p}$_B$ collinear with the 3-momentum transfer {\bf q} and opposed to it, and has the final-state 3-momentum {\bf p}$^\prime_B$ in the direction of {\bf q} and equal in magnitude to that of {\bf p}$_B$. This implies that the initial and final 3-momenta are $p_B=p'_B=q_B/2$ and therefore that the corresponding nucleon energies are $E_{p,B}=E'_{p,B}=\sqrt{m_N^2+q_B^2/4}$. Immediately the following identities emerge:
\begin{eqnarray}
{\bm \eta}_B &=& - {\bm \kappa}_B \nonumber \\ 
\theta_B &=& \pi \nonumber \\ 
\delta_B &=& 0 \nonumber \\ 
\delta'_B &=& - \eta_B \nonumber \\ 
\omega_B &=& 0 \nonumber \\ 
\lambda_B &=& 0 \nonumber \\ 
\kappa_B &=& \sqrt{\tau} \qquad\qquad \kappa_B/\sqrt{\tau} = 1 \nonumber \\ 
\varepsilon_B &=& \sqrt{1 + \kappa_B^2} = \sqrt{1 + \tau} , \label{eq-B-8}
\end {eqnarray}
where $\theta_B = \pi$ 
implies that $\sin \theta_B = 0$ and $\cos \theta_B = -1$. We shall also assume that ``L'' will refer to longitudinal polarization along ${\bm p}_B$, namely, along $-{\bm q}_B \sim -{\bm u}_3$, while ``S'' will refer to sideways polarization in the original $123$-frame, namely, in the $-{\bm u}_1$-direction; hence $\phi_B = 0$. This yields the following for the 4-vectors employed in this study:
\begin{eqnarray}
Q^\mu_B &=& q_B (0,0,0,1) = 2 m_N \sqrt{\tau} (0,0,0,1) \nonumber \\ 
P^\mu_B &=& (\sqrt{m_N^2+q_B^2/4},0,0,-q_B/2) = m_N (\sqrt{1+\tau},0,0,-\sqrt{\tau} ) \label{eq-B-10}  .  \label{eq-B-11} 
\end {eqnarray}
  
These results allow one to write the required responses in the Breit frame for the unpolarized cases using Eqs.~(\ref{eq-rot-60s}--\ref{eq-rot-63s}) in this work together with the identities $\kappa_B = \sqrt{\tau}$, $\delta_B=0$ from above. One immediately obtains
\begin{eqnarray}
\left[ W^L \right]_B &=& G_E^2 \nonumber \\ 
\left[ W^T \right]_B &=& 2 \tau G_M^2 \nonumber \\ 
\left[ W^{TT} \right]_B &=& \left[ W^{TL} \right]_B =0 , \label{eq-B-29} 
\end {eqnarray}
 Finally, employing Eqs.~(\ref{eq-nn50cE}--\ref{eq-nn53cE}) and the results above for the required  kinematic factors, the polarized results in the Breit frame are then
\begin{eqnarray}
\left[ W^{T'}_{pol} \right]^{incl}_{L,B} &=& 2 h^* \tau G_M^2 \nonumber \\ 
\left[ W^{TL'}_{pol} \right]^{incl}_{S,B} &=& -2\sqrt{2} h^* \sqrt{\tau} G_E G_M \nonumber \\ 
\left[ W^{T'}_{pol} \right]^{incl}_{S,B} &=& \left[ W^{TL'}_{pol} \right]^{incl}_{L,B} = 0 . \label{eq-B-36} 
\end {eqnarray}

\subsection{\protect\bigskip Expansions for Low Target Momentum \label{subsec-low}}

As discussed earlier, in some applications it is advantageous to expand for small $\eta$ (and hence $\delta$ and $\delta'$), retaining contributions up to order $\eta$ while dropping those of order $\eta^2$ and higher. Now, in contrast to the strict rest-frame results (see below), $\kappa$ differs from $\sqrt{\tau (1+\tau)}$ by of order $\eta$, namely
\begin{equation}
\kappa = \sqrt{\tau (1+\tau)} \left[ 1 + \sqrt{ \frac{\tau}{1+\tau} } \eta \cos\theta + {\cal O} (\eta^2) \right]. \label{eq-nn57}
\end{equation}
Using the fact that $\lambda^2 = \kappa^2 -\tau$ one then also has that
\begin{equation}
\lambda = \tau \left[ 1 + \sqrt{\frac{1+\tau}{\tau}} \eta \cos\theta + {\cal O} (\eta^2) \right]. \label{eq-nn57lam}
\end{equation}
 Expanding the general expressions in Eqs.~(\ref{eq-rot-60s}--\ref{eq-rot-63s}) for the unpolarized results gives rise to the following expanded form for the nonzero unpolarized response functions:
\begin{eqnarray}
W^L_{unpol} &=& \frac{\kappa^2}{\tau} G_E^2 + {\cal O}(\eta^2) = \frac{\kappa^2}{\tau} \left[ (1+\tau)W_2 - W_1 \right] + {\cal O}(\eta^2) \label{eq-ex-13y} \\
W^T_{unpol} &=& 2 \tau G_M^2 + {\cal O}(\eta^2) = 2 W_1 + {\cal O}(\eta^2)  \label{eq-ex-14} \\
W^{TT}_{unpol} &=& {\cal O}(\eta^2) \label{eq-ex-15} \\
W^{TL}_{unpol} &=& 2\sqrt{2} (1+\tau) W_2 \delta \cos \phi + {\cal O}(\eta^2) , \label{eq-ex-16} 
\end{eqnarray}
where, as usual,
\begin{eqnarray}
W_1 &=& \tau G_M^2 \nonumber \\ 
W_2 &=& \frac{1}{1+\tau} \left( G_E^2 + \tau G_M^2 \right) . \label{eq-extra2}
\end{eqnarray}
These unpolarized results were already derived in \cite{JandD}. As discussed above, all other unpolarized response functions are zero, and in particular only time-reversal even (TRE) responses are nonzero while all time-reversal odd (TRO) responses are zero, as they should be for elastic scattering.

In situations where $\eta$ is small one may also expand the polarized results in $\eta$ to find the following:
 \begin{eqnarray}
W^{TL'}_L  &=& 2\sqrt{2} h^* G_M \left[ \kappa G_E \sin\theta - \tau (1+\tau) \eta \sin\theta \cos\theta F_2 + {\cal O} (\eta^2) \right] \cos\phi  \label{eq-nn56} \\
W^{TL'}_S  &=& 2\sqrt{2} h^* G_M \left[ \kappa G_E \cos\theta + \tau (1+\tau) \eta \sin^2 \theta F_2 + {\cal O} (\eta^2) \right] \cos\phi \label{eq-nn56a} \\
W^{T'}_L  &=& -2 h^* G_M \left[ \tau G_M \cos\theta - \sqrt{\tau (1+\tau)} \eta \sin^2 \theta F_1 + {\cal O} (\eta^2) \right] \label{eq-nn56b} \\
W^{T'}_S  &=& 2 h^* G_M \left[ \tau G_M \sin\theta + \sqrt{\tau (1+\tau)} \eta \sin \theta \cos\theta F_1 + {\cal O} (\eta^2) \right] . \label{eq-nn56c} 
\end {eqnarray}
Note that the way that $\lambda$ enters the full expressions for the polarized response functions is via the combination $\lambda - \tau\varepsilon$ which is of order $\eta$ (see Eq.~(\ref{eq-nnn1e})) and accordingly this combination may simply be replaced by $\kappa \eta \cos\theta$ as above. 
Furthermore, one could also expand the factors $\kappa$ in Eq.~(\ref{eq-ex-13y}) and Eqs.~(\ref{eq-nn56}--\ref{eq-nn56a}) using Eq.~(\ref{eq-nn57}),  
although retaining the factors $\kappa$ leads to very compact expressions. Below we shall explore the implications of retaining terms only up to linear in $\eta$, either as in Eq.~(\ref{eq-ex-13y}) and Eqs.~(\ref{eq-nn56}--\ref{eq-nn56a}) or when also expanding $\kappa$ to linear order in $\eta$ as above; these two approximations will only differ at ${\cal O}(\eta^2 )$. While the expanded results above are potentially useful approximations, the exact results in Eqs.~(\ref{eq-rot-60s}--\ref{eq-rot-63s}) and (\ref{eq-nn56}--\ref{eq-nn56c}) are so simple that they can be employed without expanding. 

As a special case, one may also recover results in the target rest-frame where one has $\eta_R = 0$ and hence $\delta_R = 0$. For this degenerate case where the initial-state nucleon is assumed to be at rest, we assume, as in the discussion above of the Breit frame results, that the original $123$-frame is to be the one in which to express the polarizations, with ``L'' in the 3-direction and ``S'' in the 1-direction; note that these are opposite to the directions chosen when discussing the Breit frame results. For the kinematic variables in the target rest frame one has the following:
\begin{eqnarray}
\varepsilon_R &=& 1 \nonumber \\ 
\lambda_R &=& \tau \nonumber \\ 
\kappa_R &=& \sqrt{\tau (1+ \tau)} \nonumber \\ 
\rho_R &=& 1/(1+\tau)  \nonumber \\ 
\zeta_R &=& 1 \nonumber \\ 
{\bar \varepsilon}_R &=& 1 + \tau \nonumber \\ 
\alpha_{1,R} &=& 1 \nonumber \\ 
\alpha_{2,R} &=& 1 \nonumber \\ 
f_{0,R} &=& {\tilde f}_{0,R} = 1 , \label{eq-rest-7}
\end{eqnarray}
with $\rho \equiv -Q^2/q^2$, as usual. 
From Eqs.~(\ref{eq-sn-92}) in the target rest frame one then has that
\begin{eqnarray}
{\tilde \nu}_{0,R} &=& \sqrt{1+\tau} G_E \nonumber \\ 
{\tilde \nu}'_{0,R} &=& 0 \nonumber \\ 
{\tilde\nu}_{1,R} &=& 0 \nonumber \\ 
{\tilde\nu}'_{1,R} &=& \sqrt{\tau} G_M \nonumber \\ 
{\tilde\nu}'_{2,R} &=& \sqrt{\tau} G_M \nonumber \\ 
{\tilde\nu}''_{2,R} &=& 0 , \label{eq-rest-13} 
\end{eqnarray}
which then yields the following for the unpolarized responses in the rest system:
\begin{eqnarray}
\left[ W^L \right]_R &=& (1+\tau) G_E^2 = \frac{1}{\rho_R} \left( \frac{1}{\rho_R} W_2 - W_1 \right) \nonumber \\ 
\left[ W^T \right]_R &=& 2W_1 \nonumber \\ 
\left[ W^{TT} \right]_R &=& \left[ W^{TL} \right]_R =0 . \label{eq-rot-62sR}
\end{eqnarray}
Moreover, using the above expressions for the kinematic variables in the rest system (see Eqs.~(\ref{eq-rest-7},\ref{eq-rest-13})) plus the fact that 
$\lambda - \tau \varepsilon = 0$ in the rest system, and thus obtaining simplified results for the general expressions in Eqs.~(\ref{eq-nn50cE}--\ref{eq-nn53cE}), we have the following for the polarized responses in the rest system:
\begin{eqnarray}
\left[ W^{T'} \right]_R &=& - 2 h^* \tau G_M^2 \cos \theta^*_R \nonumber \\ 
\left[ W^{TL'} \right]_R  &=&  2\sqrt{2} h^* \sqrt{\tau(1+\tau)} G_E G_M \sin \theta^*_R \cos \phi^*_R \label{eq-ab26} ,
\end {eqnarray}
which are familiar results (for general discussion of polarized electron-nucleon elastic scattering see, for instance, \cite{Donnelly:1985ry,Crawford} and references therein, as well as Chapter~8 in 
\cite{thebook}). Finally, we note that these rest-frame results emerge immediately from those in the Breit frame obtained above once the directions are adjusted appropriately; namely, here the polarization angles are in the original $123$-system, whereas the Breit-frame conventions chosen above have L referring to the $-{\bm u}_3$-direction and S referring to the $-{\bm u}_1$-direction, which explains the sign reversals. Of course, one should remember that the angles in the two frames are not the same, since they are frame-dependent.

\section{\protect\bigskip Results for Unpolarized and Polarized Scattering \label{sec-num}}

While it is straightforward to employ the exact on-shell single-nucleon responses developed above, since in their final forms they are relatively compact, nevertheless it proves useful in some circumstances to explore quantitatively what happens when the expanded expressions are employed. In particular, the so-called ``prescription for nuclear physics'' for studies of electron scattering from nuclei (see also below) entails taking the on-shell single-nucleon momentum space current operators as developed here and using some approximation to them to deduce effective single-nucleon operators for use in coordinate space. If one uses the Fermi momentum $k_F$ to gauge the scale of the missing momentum below which the PWIA cross sections are relatively large then the corresponding dimensionless scale is $\eta_F \equiv k_F/m_N$. For all but the few-body nuclei this scale does not change radically across the periodic table (see \cite{Maieron} for determinations of the relevant values of $k_F$) --- these vary from $k_F (^{12}$C$) = 228$ MeV/c to $k_F (^{208}$Pb$) = 248$ MeV/c, and correspondingly one has $\eta_F (^{12}$C$) = 0.25$ and $\eta_F (^{208}$Pb$) = 0.26$. Accordingly, for many nuclei across the periodic table we expect that when $\eta$ lies below $\sim$1/4 the PWIA cross sections are relatively large. The few-body nuclei are somewhat different: the values of the Fermi momenta in that case are $k_F (^{2}$H$) = 55$ MeV/c, $k_F (^{3}$He$) = 150$ MeV/c and $k_F (^{4}$He$) = 200$ MeV/c, which yield $\eta_F (^{2}$H$) = 0.06$, $\eta_F (^{3}$He$) = 0.16$ and $\eta_F (^{4}$He$) = 0.21$. However, the few-body nuclei have spectral functions that peak at zero missing momentum, being dominated by S-wave components in their ground states, and thus the relevant values of missing momenta for these systems are even smaller than the Fermi momentum estimates given here. 

Thus, if one chooses to work only to leading order in $\eta$, neglecting contributions of ${\cal O}(\eta)$ and higher, then for typical nuclei one may incur errors of order 25\% in the regime where the PWIA cross sections are large, with smaller errors for the few-body nuclei. However, if one includes the next order, namely if one includes terms of ${\cal O}(\eta)$, but neglects terms of ${\cal O}(\eta^2)$ as discussed above, then the errors are reduced to only a few percent.

If one attempts to go to very large values of missing momentum then these arguments become invalid; however, in such a case it is less likely that the strategy of using on-shell single-nucleon currents is a robust approximation.

Next we consider the single-nucleon electron scattering responses where we choose to express these for a specific choice of a complete set of the input kinematic variables.  Namely all results are given for $\phi = 0$, since the dependence on this variable is very simply through the factors of $\cos\phi$ or $\cos 2\phi$ in Eqs.~(\ref{eq-rot-62s}), (\ref{eq-rot-63s}), 
(\ref{eq-nn50cE}) and (\ref{eq-nn51bE}). We fix $\tau$, which has the advantage of fixing the values of the electromagnetic form factors, and for chosen values of the angle $\theta$ show results as functions of the target 3-momentum $p$ or correspondingly its dimensionless version $\eta$.


Now, we provide numerical results for the responses. We compare the ``full results'' in Eqs.~(\ref{eq-rot-60s}--\ref{eq-rot-63s}) and (\ref{eq-nn50cE}--\ref{eq-nn53cE}) with two specific first-order approximations: one is called the ``single first-order approximation'' wherein factors involving the ratio $\kappa/\sqrt{\tau}$ are not approximated, but where higher-order contributions in $\eta$ to the responses are neglected, and a second called the ``double first-order approximation'' in which the factors involving $\kappa/\sqrt{\tau}$ are also expanded to linear order in $\eta$. Specifically, for the unpolarized responses this implies that the single first-order responses are taken to be the following
\begin{eqnarray}
\left[ W^L \right]_{single} &=& \frac{\kappa^2}{\tau} G_E^2 \nonumber \\ 
\left[ W^T \right]_{single}  &=& 2W_1  \nonumber \\ 
\left[ W^{TT} \right]_{single} &=&  0 \nonumber \\ 
\left[ W^{TL} \right]_{single} &=& 2\sqrt{2} ( 1 + \tau )\delta W_2  \cos \phi , \label{eq-rot-63s-1}
\end{eqnarray}
while the double first-order responses also employ Eq.~(\ref{eq-nn57-z}) where $\kappa/\sqrt{\tau}$ is expanded to first order in $\eta$; only $W^L$ is affected. For the polarized responses we use Eqs.~(\ref{eq-nn56}--\ref{eq-nn56c}) expanding in $\eta$, namely, for the single first-order approximation we have
 \begin{eqnarray}
\left[ W^{TL'}_L \right]_{single}  &=& 2\sqrt{2} h^* G_M \left[ \kappa G_E \sin\theta - \tau (1+\tau) \eta \sin\theta \cos\theta F_2 \right] \cos\phi  \nonumber \\ 
\left[ W^{TL'}_S \right]_{single}  &=& 2\sqrt{2} h^* G_M \left[ \kappa G_E \cos\theta + \tau (1+\tau) \eta \sin^2 \theta F_2 \right] \cos\phi \nonumber \\ 
\left[ W^{T'}_L \right]_{single}  &=& -2 h^* G_M \left[ \tau G_M \cos\theta - \sqrt{\tau (1+\tau)} \eta \sin^2 \theta F_1 \right] \nonumber \\ 
\left[ W^{T'}_S \right]_{single}  &=& 2 h^* G_M \left[ \tau G_M \sin\theta + \sqrt{\tau (1+\tau)} \eta \sin \theta \cos\theta F_1 \right] , \label{eq-nn56c-1} 
\end {eqnarray}
while for the double first-order approximation we also expand $\kappa$ to linear order in $\eta$ again using Eq.~(\ref{eq-nn57-z}); only the $TL'$ responses are affected. 
For comparison, the rest frame answers are given in Eqs.~(\ref{eq-rot-62sR}) and Eqs.~(\ref{eq-ab26})

We fix the value for $\tau$, so that the 
values for the nucleon form factors are also fixed. We focus on the proton case, although later we also comment on what differences are to be found when neutrons are considered or, alternatively, when the focus is placed on isoscalar/isovector form factors. We use the dipole parametrization for the proton form factors for simplicity, although none of our results is particularly sensitive to the choice we make here. 

Then, we fix the polar angle $\theta$ and use $\eta$ as a variable. The azimuthal angle $\phi$ is set to zero here --- the only place where it shows up is in the form of overall factors of $\cos \phi$ and $\cos 2 \phi$ for the interference responses. We choose to consider only values of $\eta < 0.6$ which, as discussed previously and  below, is actually quite large for most nuclear physics applications. In order to keep the number of figures manageable, we will present results for the values of $\theta = 0^o, 180^o$ only for the responses that have non-trivial values there, and then the results for $\theta = 45^o, 90^o$. We will present both the values for the responses, which allows one to compare the size of the different responses directly, and give some numerical values for the ratios of the single and double first-order responses to the full response results. The latter graphs will show most clearly when approximations in lowest order are acceptable, and when they lead to major deviations. 
 
\begin{figure}
	\centering
	\includegraphics[width=14pc]{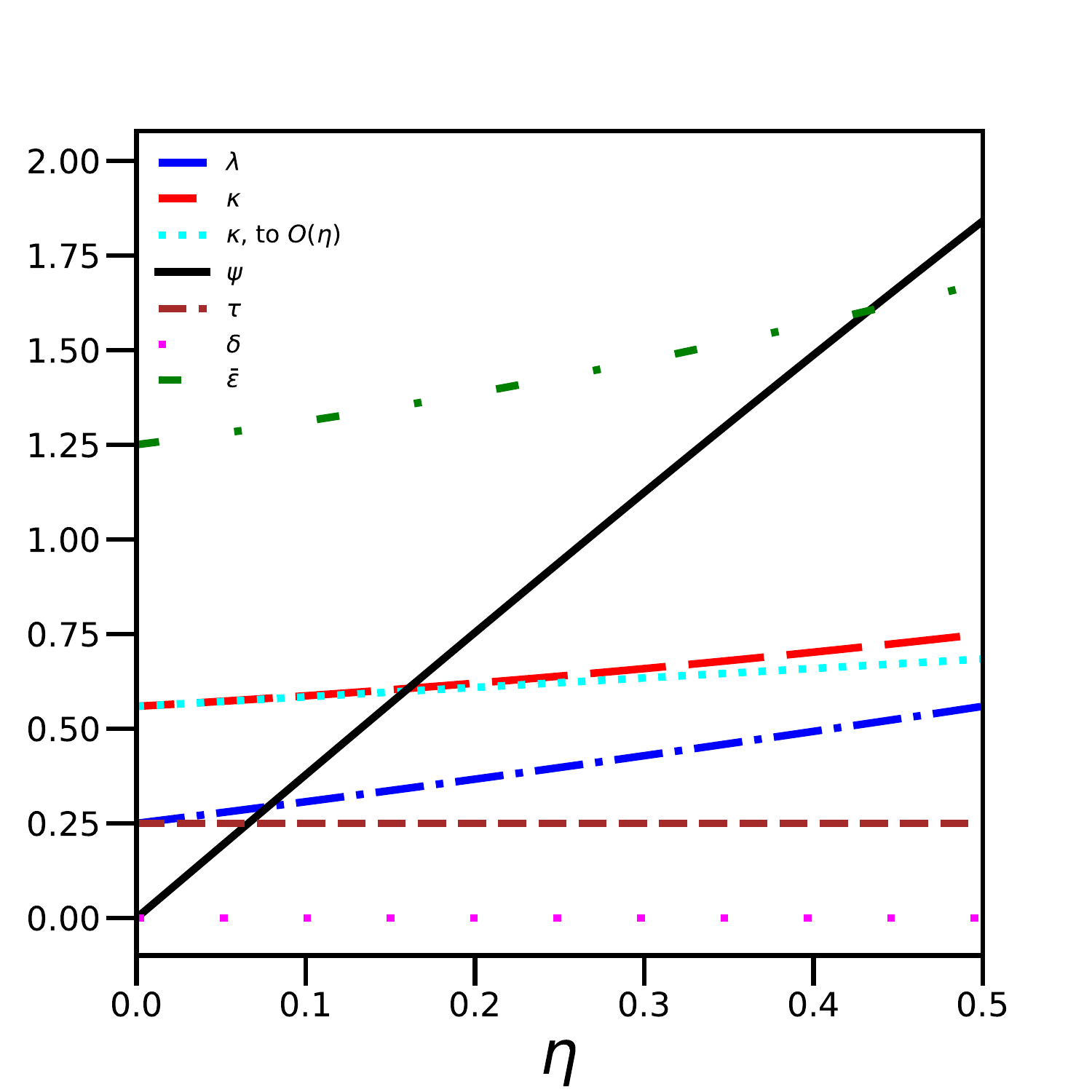} 
	\includegraphics[width=14pc]{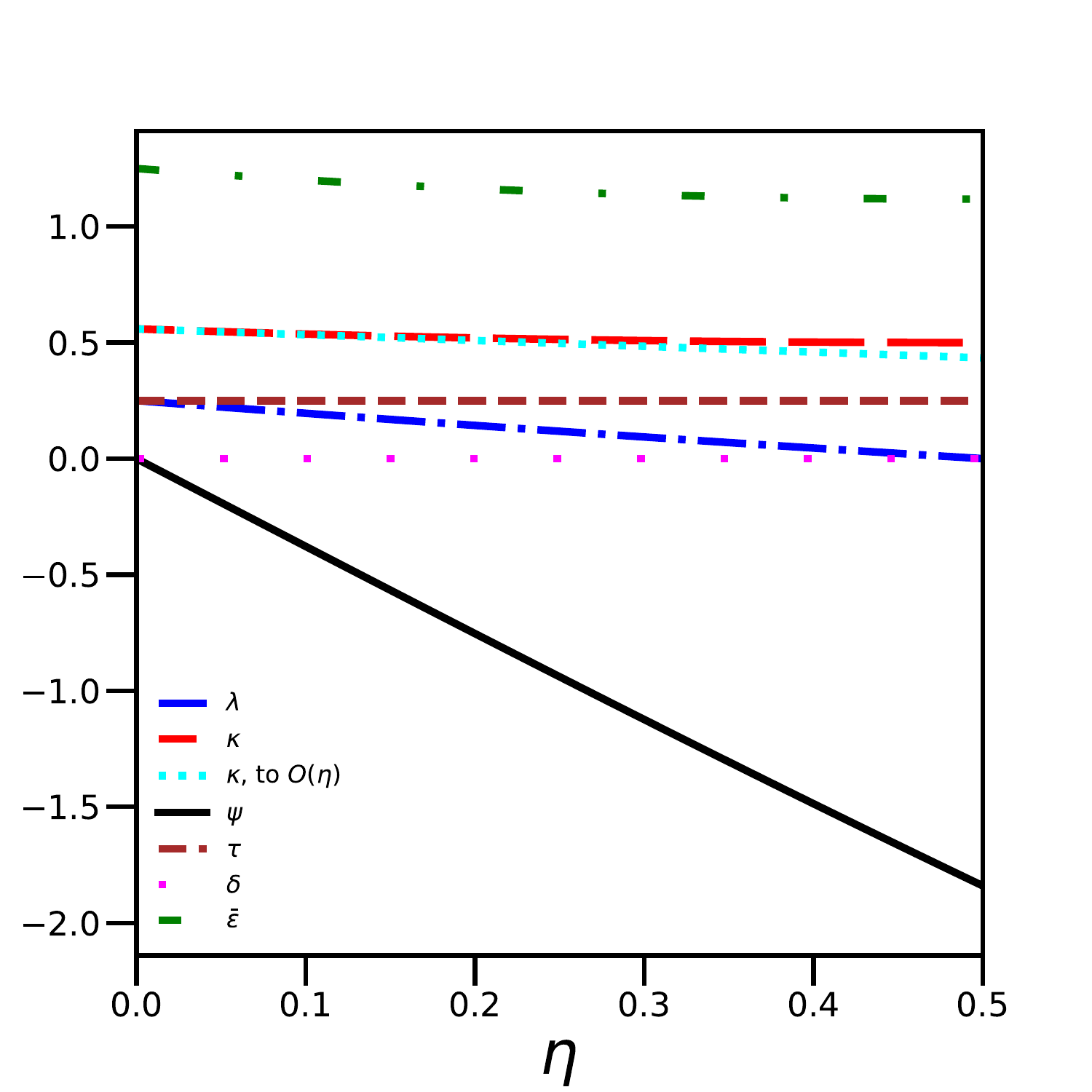}
	 \includegraphics[width=14pc]{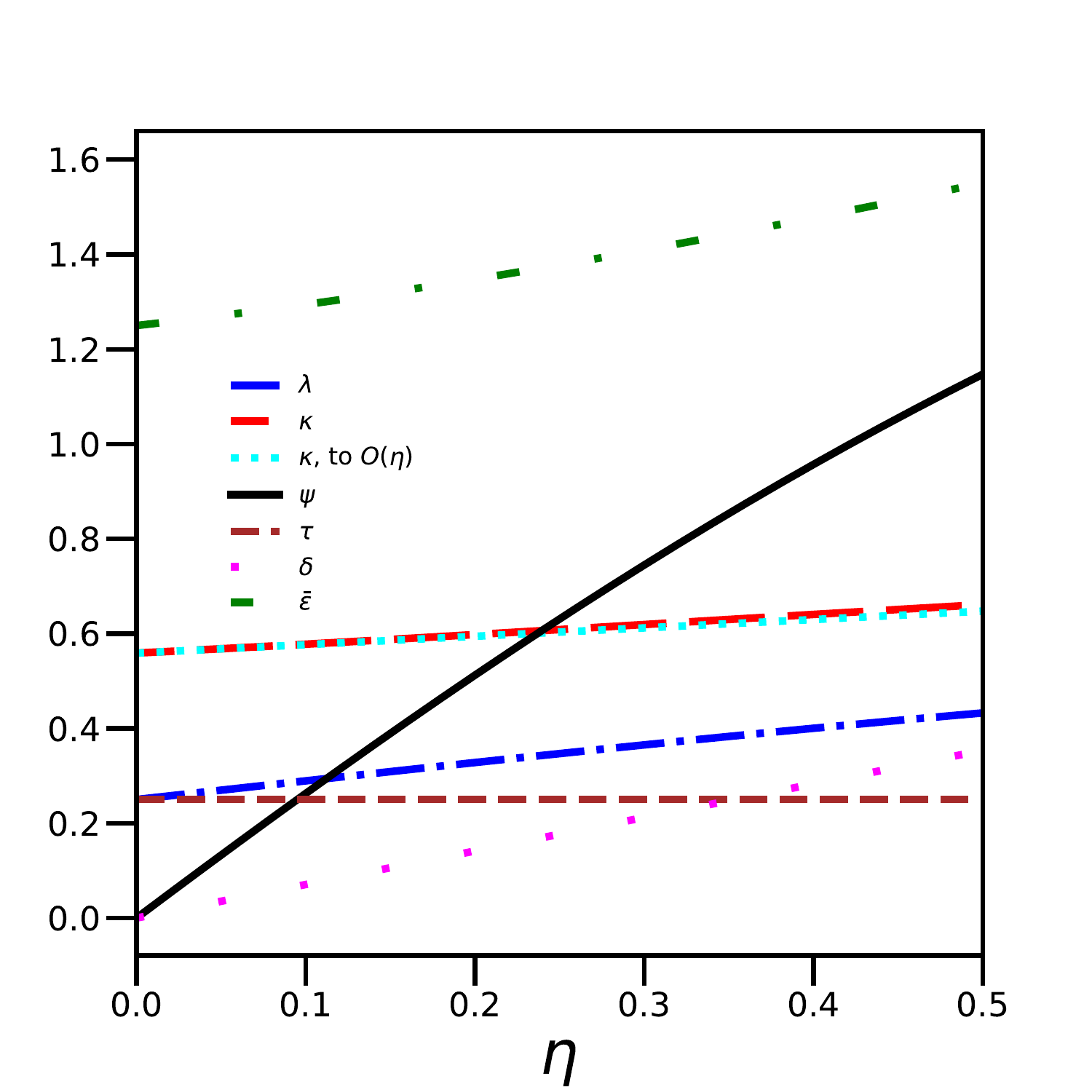} 
	 \includegraphics[width=14pc]{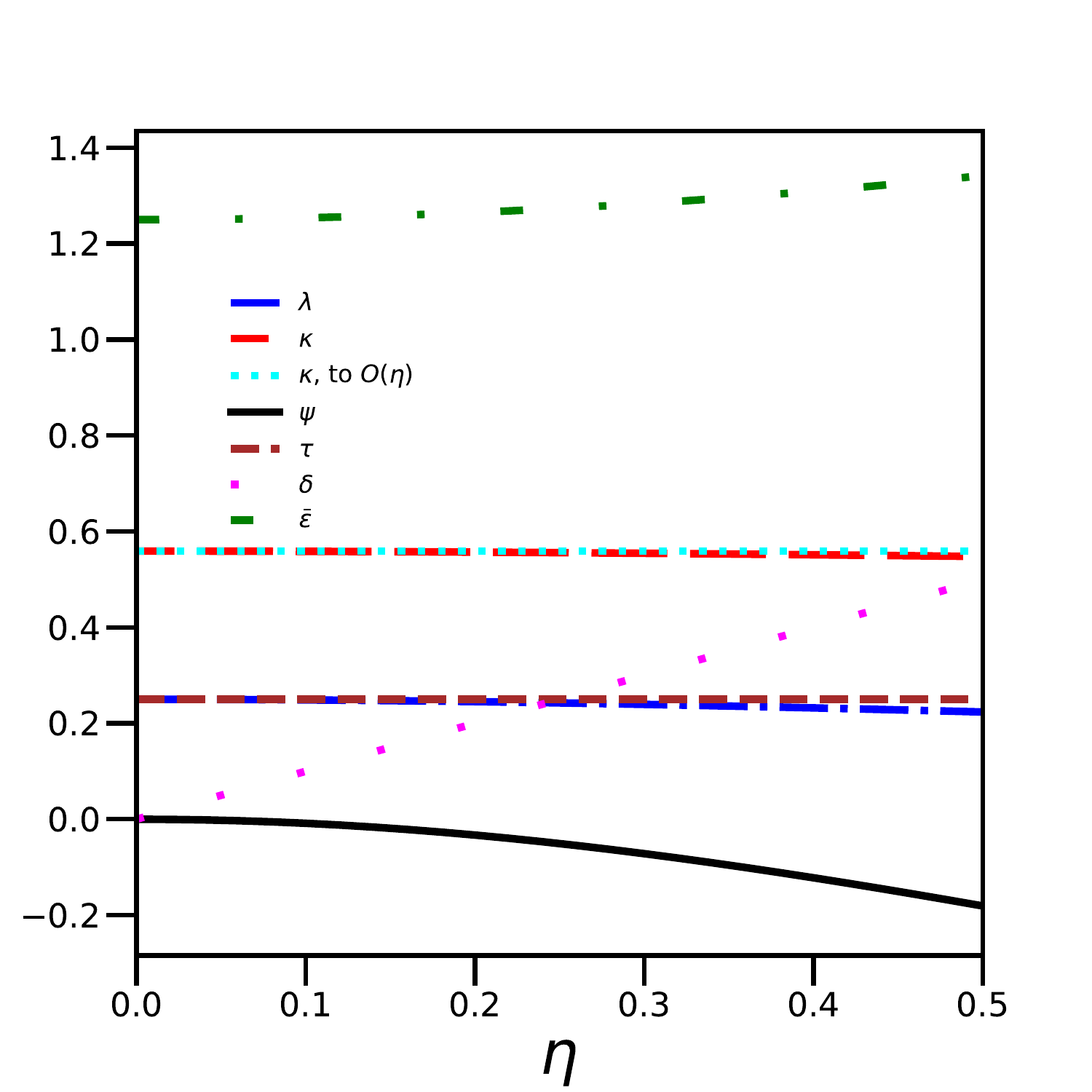}
	\caption{Various dimensionless kinematic variables are shown as functions of $\eta$ for $\tau = 0.25$ and
		$\theta = 0^o$ (top left  panel), $\theta = 180^o$ (top right panel), 
			$\theta = 45^o$ (bottom left  panel), and $\theta = 90^o$ (bottom right panel). For each panel,
			the black solid line is $\psi$, the blue long-dash-dotted line is $\lambda$, the red long-dashed line is $\kappa$, the cyan dotted line is $\kappa$ to $O(\eta)$, the brown dashed line is $\tau$,
			the red loosely dotted line is $\delta$, and the green loosely dash-dotted line is $\bar{\epsilon}$.
		 }
	\label{fig-kinematicsall}
\end{figure}

We will start our presentation of numerical results with a value of $\tau = 0.25$, namely, for $|Q^2| \sim 1$ (GeV/c)$^2$. By fixing $\tau$ and therefore $Q^2$, we deal with the same value for the form factors
throughout the kinematics. We have chosen $\eta$ as independent variable and fix the value of the angle $\theta$
for each panel with numerical results, using $\theta = 0^o, 45^o, 90^o, 180^o$, thus scanning the entire
phase space. Various kinematics variables are plotted in Fig. \ref{fig-kinematicsall}. Among them are the 
exact value for $\kappa$, and the first order in $\eta$ approximation to $\kappa$, using Eq.~(\ref{eq-nn57-z}).
It is clear from all four panels, showing results for  $\theta = 0^o, 45^o, 90^o, 180^o$ that the full value of $\kappa$ and the first-order approximation are numerically quite close. The biggest deviations are observed for
larger $\eta$ values for $\theta =0^o$ and $\theta = 180^o$.

We also include the scaling variable $\psi$ introduced in \cite{Alberico:1988bv},
\begin{equation}
	\psi = \frac{1}{\sqrt{\xi_F}} \frac{\lambda - \tau}{\sqrt{(1 + \lambda) \tau + \kappa \sqrt{\tau (\tau+ 1)}}} , \label{eq-psi}
\end{equation}
where 
\begin{eqnarray}
	\epsilon_F & \equiv& \sqrt{1 + \eta_F^2} \nonumber \\ 
	\xi_F & \equiv& \epsilon_F - 1 \nonumber \\ 
	\eta_F & \equiv & k_F/m_N \label{eq-psi-3} . 
\end{eqnarray}
In order to calculate $\psi$, we have employed a value of $k_F = 0.25$ GeV/c \cite{Maieron} as being representative of medium to heavy nuclei. 

As discussed in \cite{Alberico:1988bv} the Relativistic Fermi Gas (RFG) model for the nuclear EM response is based directly on the fully covariant single-nucleon matrix elements that constitute the focus of the present work. In the non-Pauli-blocked region the RFG cross section may be written in terms of this variable, where then the quasielastic peak occurs at $\psi = 0$ and the response falls to zero when $\psi = \pm 1$. We take this model only as a rough measure of where the nuclear quasielastic response is large, say in the region $-0.7 < \psi < +0.7$, which we denote as the ``Fermi cone'' and outside this region which we call the ``tail regions''. Accordingly, in evaluating the results to follow one should distinguish what happens in the Fermi cone where strength may be assumed to be significant (at least if the RFG is used as an indicator) from the tail regions where typically the nuclear response is smaller. 

When considering the kinematics variables plotted in Fig. \ref{fig-kinematicsall}, it is obvious that $\psi$ changes considerably with $\eta$ for $\theta = 0^o, 180^o$. For these two angles $\theta$, the values for $\psi$ correspond to the Fermi cone roughly for $\eta < 0.2$, indicating that we would probe both
regions of large response (Fermi cone) and small response values (Fermi tail), when scanning the whole range of $\eta$ values for applications in medium or large nuclei. For $\theta = 45^o$, the change-over between Fermi cone and Fermi tail is happening around $\eta \approx 0.3$, with lower $\eta$ values corresponding to the
Fermi cone. Interestingly, for $\theta = 90^o$, the whole range of $\eta$ leads to values of $|\psi|$ considerably less than 0.7, so the only region relevant
for these kinematics is the Fermi cone.  We are considering these regions in $\psi$ so that we have an idea where to expect larger cross sections when applying our
results for the electron-nucleon cross section to electron-nucleus scattering, and to get a handle on the practical relevance of 
differences in approximations to the electron-nucleon responses for the nuclear case.

\begin{figure}
	\centering
	\includegraphics[width=16cm]{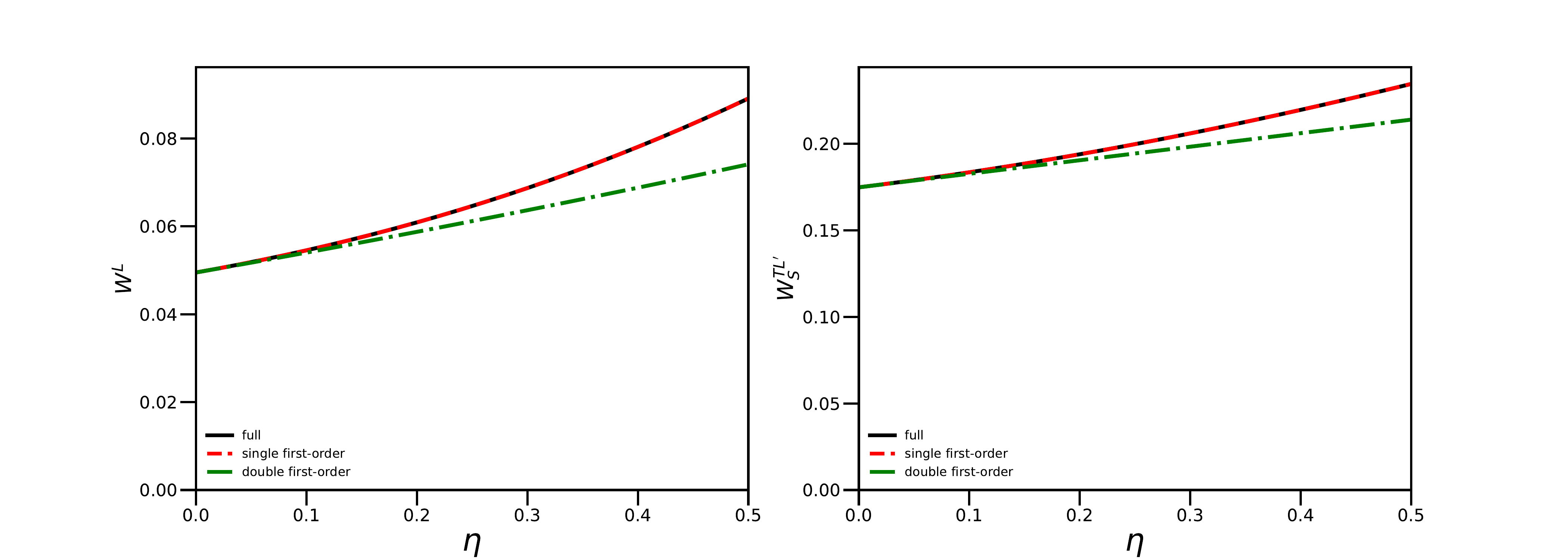} 	
	\includegraphics[width=16cm]{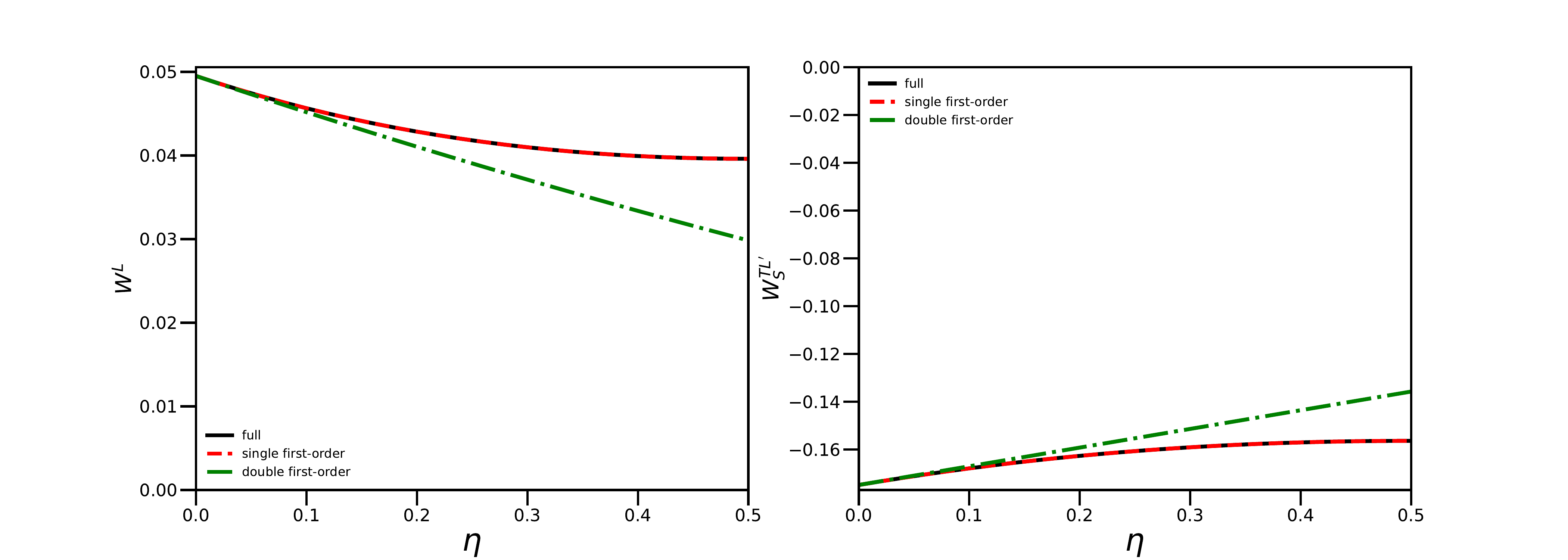} 	
	\caption{The two responses $W^L$ (left panels) and $W^{TL'}_S$ (right panels) are shown as functions of $\eta$ for $\tau = 0.25$ and
		$\theta = 0^o$ (top panels) and $\theta = 180^o$ (bottom panels). For these two $\theta$ values,  $W^{TT} = W^{TL} = W^{TL'}_L = W^{T'}_S  = 0$, and  $W^T$ and $W^{T'}_L$ reduce to
		the value of $ 2 W_1 = 0.154$ and $\pm 2 h^* W_1$, respectively.}
	\label{fig-resptheta0180}
\end{figure}

We look first at the results for $\theta = 0^o$ and $\theta = 180^o$, as the only
two interesting responses at these values of $\theta$ are $W^L$ and $W^{TL'}_S$. For these two $\theta$ values, the two responses $W^T$ and $W^{T'}_L$ reduce to
the value of $ 2 W_1$ and $\pm 2 h^* W_1$, respectively, for the full and both single first-order and double first-order approximations. The value of $2 W_1$ happens to be just above 0.15 for $\tau = 0.25$. The remaining four responses,
$W^{TT}, W^{TL}, W^{TL'}_L$ and $W^{T'}_S$ are equal to zero for $\theta = 0^o$ and $\theta = 180^o$.
In Fig. \ref{fig-resptheta0180}, we show the two relevant responses. The longitudinal response shows agreement, with the value given by $(\kappa^2/\tau) G_E^2$, between the full version and the single first-order approximation in $\eta$. 
The double first-order result for both $\eta$ and $\kappa$ starts to deviate from the full result for $\eta > 0.1$, {\it i.e.,} already in the region belonging to the Fermi cone, due to the deviation of the lowest order approximation of $\kappa$ from the exact value of $\kappa$. For the highest values of $\eta$ we consider, the ratio for $W^L$, in the double first-order approximation, dips down close to $70 \%$, and 
the ratio for $W^{TL'}_S$ holds up much better, only reaching $80 \%$.
In Fig. \ref{fig-resptheta0180}, for $\theta = 0^o$, the full result and single first-order approximation are the largest for both responses. For $\theta = 180^o$, the ordering of the different approximations is different
for $W^{TL'}_S$, due to the different behavior of $\kappa$ for the different angles, see Fig.~\ref{fig-kinematicsall}. Here,  the double first-order results are smallest. For $\theta = 0^o$, $\kappa$ increases slightly with increasing $\eta$, whereas for $\theta = 180^o$, $\kappa$ decreases slightly with increasing $\eta$. 
We now consider the numerical results for $W^{TL'}_S$ at the two $\theta$ values. The behavior of the different approximations at $\theta = 0^o$ is qualitatively very similar to $W^L$: the full and single first-order  approximation agree, and are the largest. The double first-order approximation in both $\eta$ and $\kappa$  deviates noticeably from the full result for $\eta > 0.1$. 
For $\theta = 180^o$, the sign of $W^{TL'}_S$ changes due to the factor of $\cos \theta$ in the non-vanishing part of the response. And just like for $W^L$, the ordering of the sizes is different, with the double first-order approximation having the smallest magnitude. The size and onset of deviations from the full result is the same as for $\theta = 0^o$.

\begin{figure}
	\centering 
	\includegraphics[width=16cm]{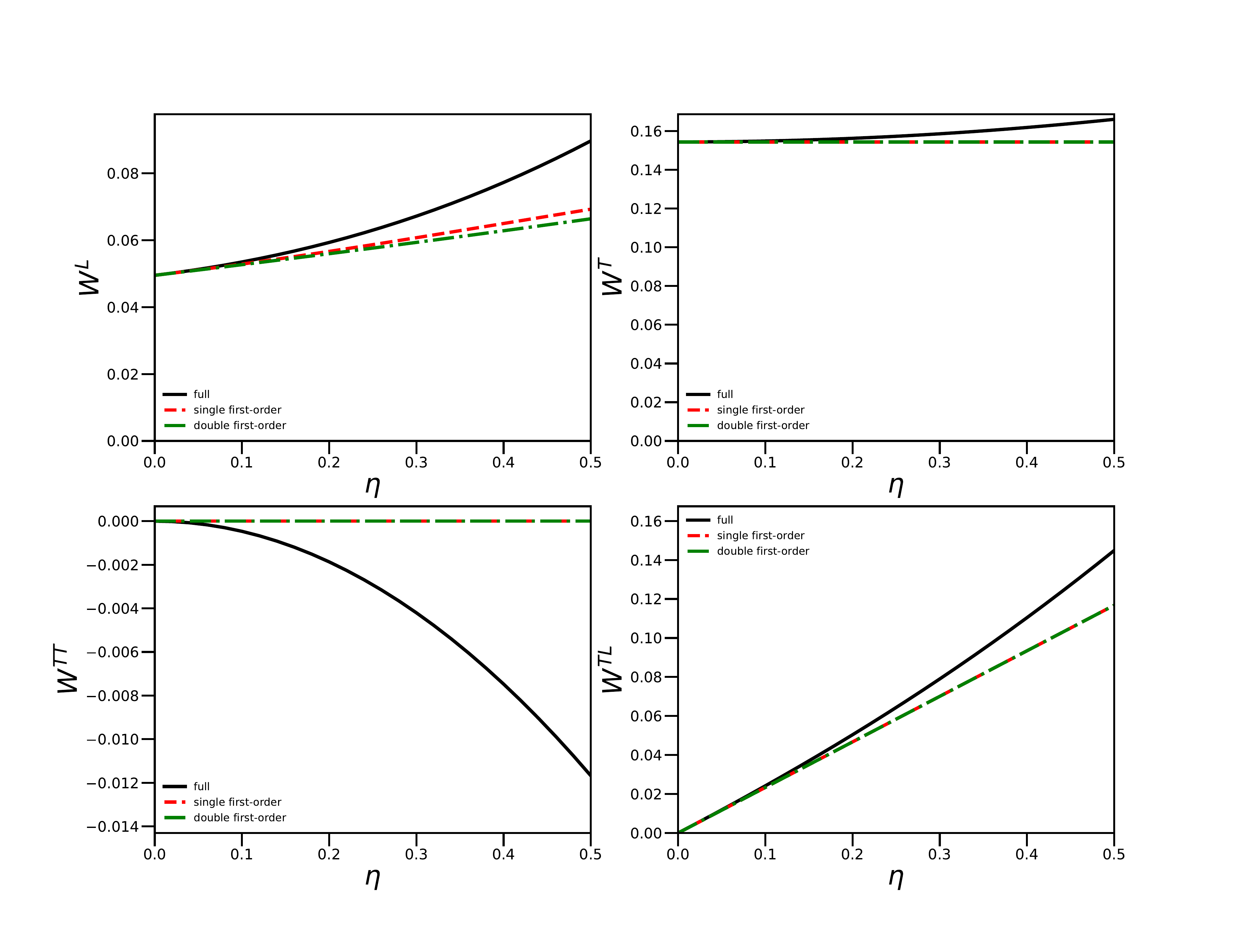} 				
	\caption{The unpolarized responses are shown as functions of $\eta$ for $\tau = 0.25$ and
		$\theta = 45^o$. The top row shows $W^L$ (left) and $ W^T$ (right), and the bottom row shows
		$W^{TT}$ (left) and $W^{TL}$ (right).
		 The solid line in these four panels shows the full solution, the red dashed line the single first-order approximation and the dash-dotted green line the double first-order approximation. }
	\label{fig-respunpoltheta45}
\end{figure}

\begin{figure}
	\centering
	\includegraphics[width=16cm]{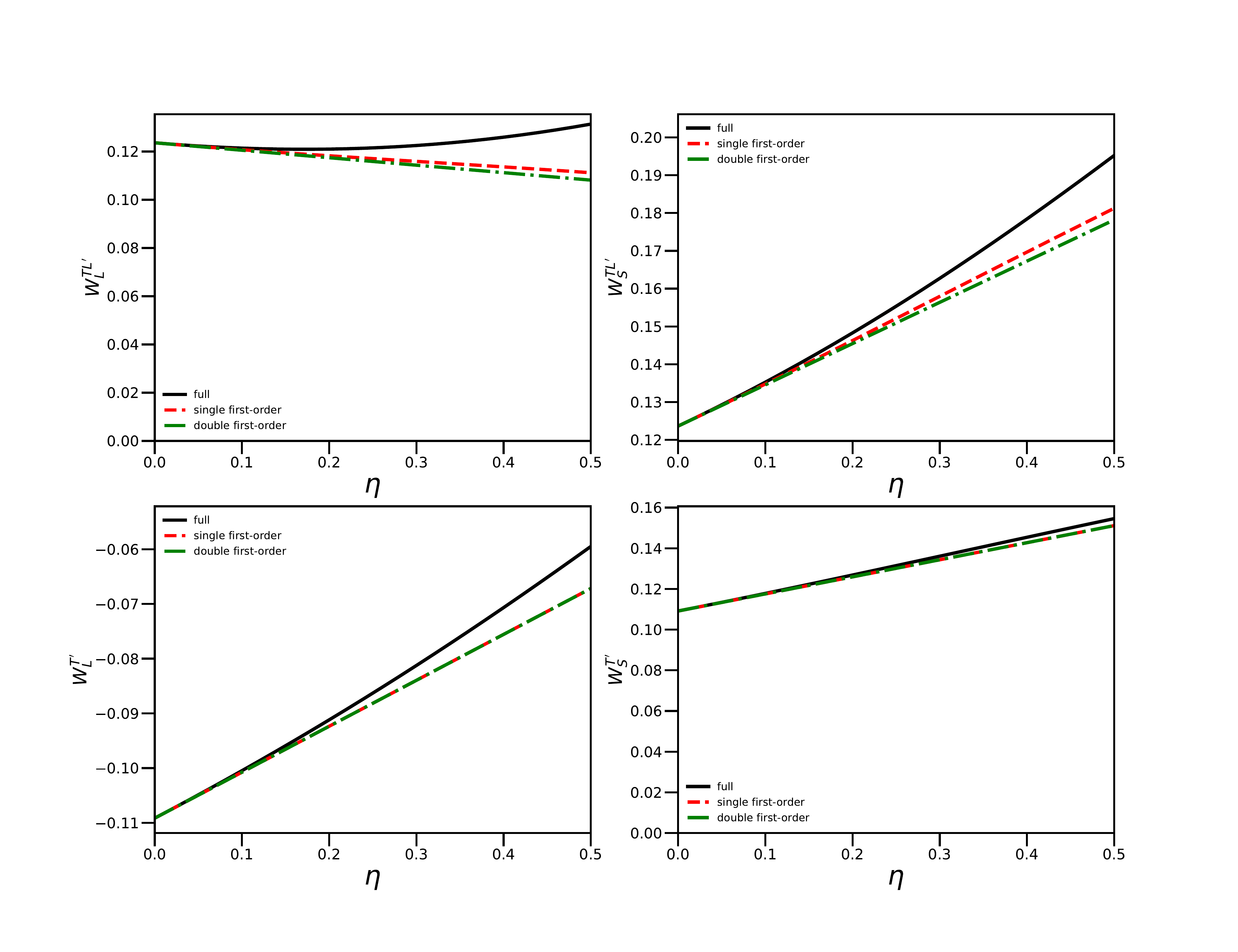} 				
	\caption{The polarized responses are shown as functions of $\eta$ for $\tau = 0.25$ and
		$\theta = 45^o$. The top row shows $W^{TL'}_L$ (left) and $ W^{TL'}_S$ (right), and the bottom row shows
		$W^{T'}_L$ (left) and $W^{TL'}_S$ (right). The solid line in these four panels shows the full solution, the red dashed line the single first-order approximation and the dash-dotted green line the double first-order approximation. }
	\label{fig-resppoltheta45}
\end{figure}

Now we consider the angles $\theta = 45^o, 90^o, 135^o$. Here, all unpolarized and polarized responses show interesting behavior. We will start with $\theta = 45^o$ and the unpolarized responses, see Fig. \ref{fig-respunpoltheta45}.
For the longitudinal response $W^L$, we now observe a difference between the full solution and the single first-order approximation in $\eta$, due to the $\delta^2 W_2$ term in the full expression: $\delta = \eta \sin \theta$ is non-zero for
$\theta \not = 0^o, 180^o$ and thus leads to a noticeable difference between the two expressions. The two first-order expressions also differ slightly from each other, but are quite close throughout. At $\eta = 0.2$, roughly the changeover between the Fermi cone and tail regions, the ratio of the single first-order approximation to the full result is equal to $95.5\%$, and for the
double first-order approximation, the corresponding ratio is $94.3\%$. For $\eta = 0.5$, these ratios take the
values of $77.2\%$ and $74.0\%$. 
For the transverse response $W^T$, the term $\delta^2 W_2$ now distinguishes the full response from both approximations, and the differences increase with increasing $\eta$, just like for $W^L$. The first-order approximations are all identical, taking the rest value $2 W_1$. However, throughout the whole range of $\eta$ for which we 
calculate the responses, the percentage deviation is quite small: at $\eta = 0.2$, the ratio of the single first-order approximation to the full result is equal to $98.8\%$, and the ratio of the double first-order approximation is exactly the same. For $\eta = 0.5$, both of these ratios take the
value of $93.0\%$.
For $W^{TT}$, only the full expression is nonzero. The magnitude of the response increases with $\eta$, as it is proportional to $\delta^2$.
For $W^{TL}$, the two first-order results coincide. The full result is just a bit above the first-order values, and the difference increases with $\eta$, and becomes noticeable only for $\eta > 0.2$,  {\it i.e.,} in the Fermi cone region. At $\eta = 0.2$, the ratio of either first-order approximation to the full result is equal to $92.7\%$, and at $\eta = 0.5$, the corresponding ratio is $80.6\%$. The numerical difference between full and first-order expressions stems from the
difference between $\bar{\varepsilon}$, which increases slightly with $\eta$, see the bottom left panel of Fig. \ref{fig-kinematicsall}, and the factor $ 1 + \tau$, which we hold constant.

We now present the polarized responses for $\theta = 45^o$ in Fig. \ref{fig-resppoltheta45}.
For the polarized responses, we observe a similar behavior to $W^L$ in both $W^{TL'}_L$ and $W^{TL'}_S$. The full result leads to the largest values, 
but below $\eta = 0.2$, corresponding to the Fermi cone, the difference to the first-order approximations
is very small. At $\eta = 0.2$, the ratio of the single first-order approximation to the full result for $W^{TL'}_L$ is $97.7\%$, and for the double first-order approximation, the ratio to the full result is $97.0\%$.
For $W^{TL'}_S$, these ratios are $98.6\%$ and $98.1\%$, respectively.
 For larger values of $\eta$, the gap between the full solution and the two approximations widens.
Both first-order approximations remain very close to each other, the small difference stems from the
factor $\kappa$ multiplying the $G_E$ term in Eqs.~(\ref{eq-nn50cE}), (\ref{eq-nn51bE}). At $\eta = 0.5$, the ratio of single first-order to full result for $W^{TL'}_L$ drops to $84.6\%$, and the double first-order to full ratio is $82.3\%$. For $W^{TL'}_S$, these ratios are $92.8\%$ and $91.3\%$, respectively.
For the polarized $W^{T'}$ responses, the two first-order approximations agree throughout, as there is no
$\kappa$ in the expressions for the first-order approximation for $W^{T'}$ in Eqs.~(\ref{eq-nn52bE}), (\ref{eq-nn53cE}). For $W^{T'}_L$,
the first-order results deviate for $\eta > 0.2$, still in the Fermi cone region for this $\theta$ value, from the full result. At $\eta = 0.2$, the ratio of first-order to full result is $101\%$, and at $\eta = 0.5$, it is $113 \%$.
%
For $W^{T'}_S$, full and both first-order approximations are extremely close to each other for the entire range of $\eta$ considered. At $\eta = 0.2$, the ratio of first-order to full result is $99.3\%$, and at $\eta = 0.5$, it is $97.8 \%$. 

For $\theta = 135^o$,  we qualitatively observe the same trends as for $\theta = 45^o$. The shape of 
some of the polarized responses changes (increasing or decreasing) due to the way the various kinematic quantities change with the angle $\theta$, as well as the signs of the cosine functions. For brevity, we omit showing this figure.

\begin{figure}
	\centering
	\includegraphics[width=16cm]{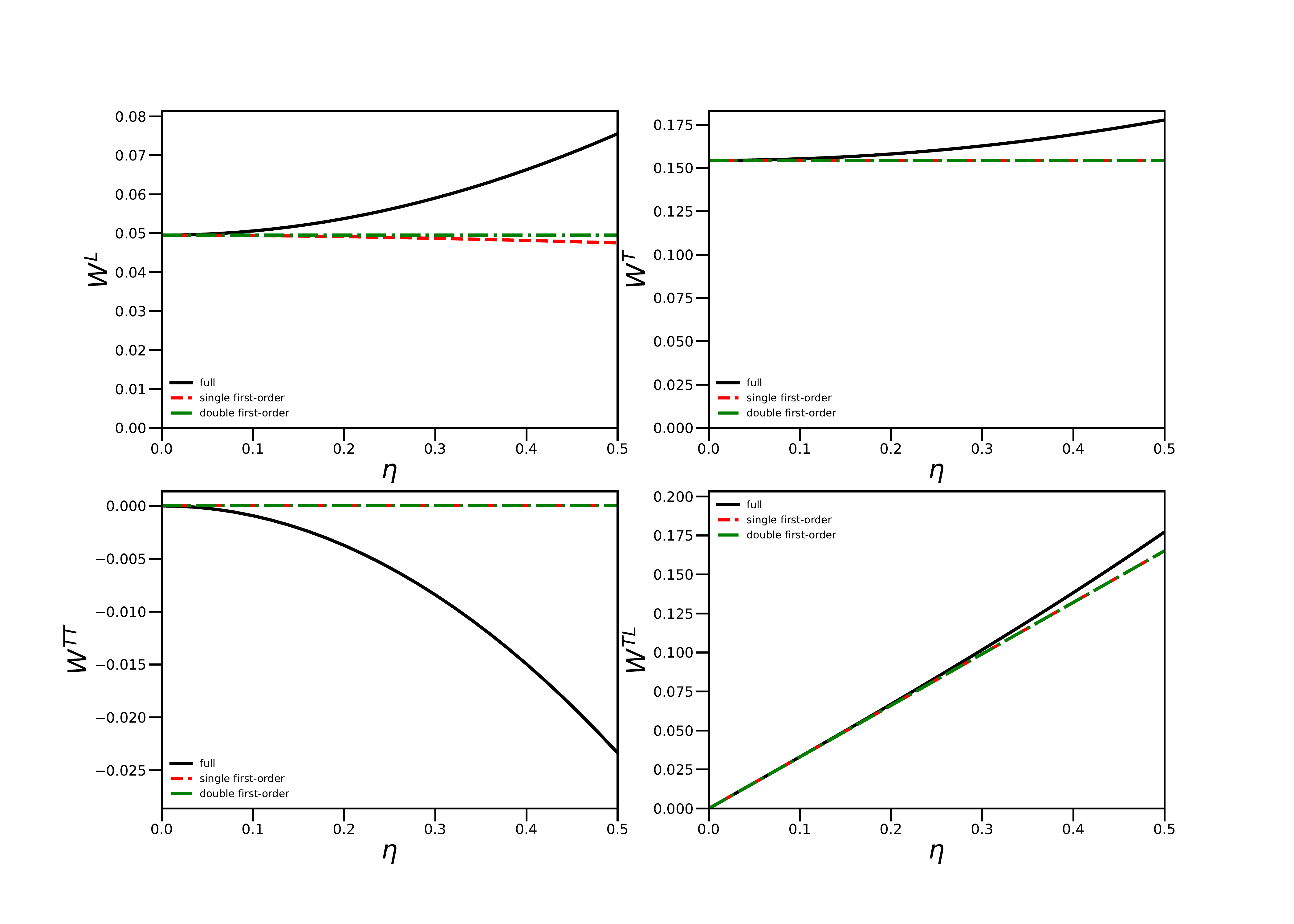} 				
	\caption{The unpolarized responses are shown as functions of $\eta$ for $\tau = 0.25$ and
		$\theta = 90^o$. The top row shows $W^L$ (left) and $ W^T$ (right), and the bottom row shows
		$W^{TT}$ (left) and $W^{TL}$ (right).
		The solid line in these four panels shows the full solution, the red dashed line the single first-order approximation and the dash-dotted green line the double first-order approximation. }
	\label{fig-respunpoltheta90}
\end{figure}

\begin{figure}
	\centering
	\includegraphics[width=16cm]{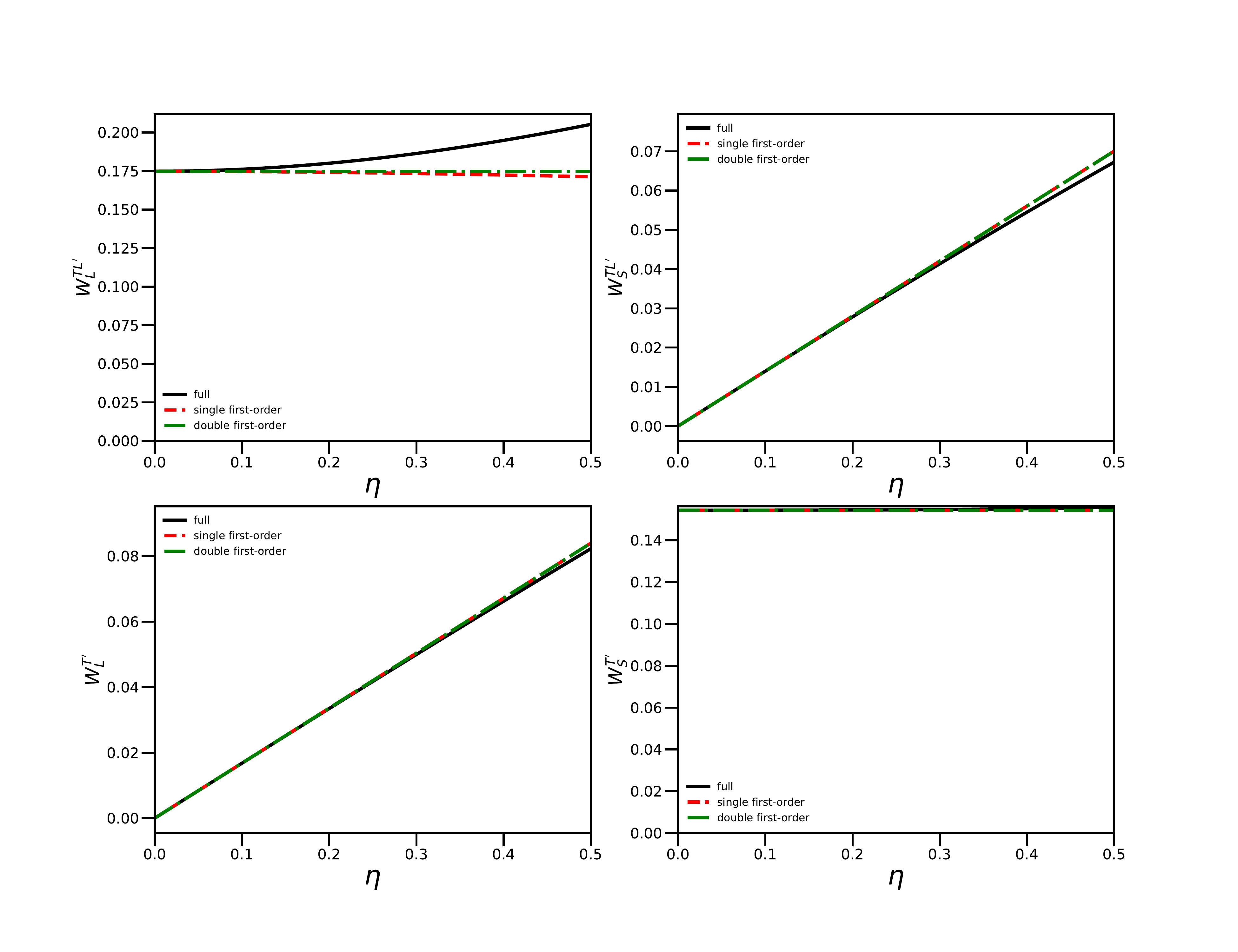} 				
	\caption{The polarized responses are shown as functions of $\eta$ for $\tau = 0.25$ and
		$\theta = 90^o$. The top row shows $W^{TL'}_L$ (left) and $ W^{TL'}_S$ (right), and the bottom row shows
		$W^{T'}_L$ (left) and $W^{TL'}_S$ (right). The solid line in these four panels shows the full solution, the red dashed line the single first-order approximation and the dash-dotted green line the double first-order approximation. }
	\label{fig-resppoltheta90}
\end{figure}

Finally, we take a look at $\theta = 90^o$, with the unpolarized responses shown in Fig. \ref{fig-respunpoltheta90}. For this value of $\theta$, we observe that for $W^L, W^T$ and $W^{TL'}_L$, the single first-order approximation and the double first-order approximation differ visibly from the
full expression  for $\eta > 0.2$.  Note that for $\theta = 90^o$, all values of $\eta$ correspond to the Fermi cone region, so these differences at higher $\eta$ are very significant for nuclear applications.  However, the single first-order and double first-order approximations agree (almost) perfectly with each other. There is a slight difference for $W^L$ for the single first-order and double first-order results to the $\kappa$ dependence in the first-order expression. Here, for $\theta = 90^o$, the double first-order expression is slightly larger than the single first-order approximation, and thus slightly closer to the full result. This is due to the behavior of the first-order approximation of $\kappa$, which is slightly larger than the actual, unapproximated value of $\kappa$, see the corresponding panel in Fig. \ref{fig-kinematicsall}. For $\eta = 0.2$, the ratio of the single first-order approximation to the full result is
$91.4\%$ and for the double first-order approximation, the ratio is $92.1\%$. At $\eta = 0.5$, these two ratios 
are equal to $62.9\%$ and $65.5 \%$.
For $W^T$ and $W^{TL}$, the two first-order approximations coincide, as already pointed out above. For $W^T$, they do deviate a bit more from the full result here compared with the $\theta = 45^o$ result, as expected, as the difference is $\propto \delta^2 W_2$,
and $\delta = \eta \sin \theta$ is largest for $\theta = 90^o$. For $\eta = 0.2$, the ratio of the first-order approximations to the full result is $97.6\%$, and for $\eta = 0.5$, it is $86.9\%$.
In the case of $W^{TT}$, the disagreement with the full result starts right at $\eta = 0$, as the first-order approximations are both zero. 
For $W^{TL}$ all three expressions agree for $\eta < 0.3$, and differences for higher values of $\eta$ are minimal. For $\eta = 0.2$, the ratio of the first-order approximations to the full result is $ 98.8\%$, and for $\eta = 0.5$, it is $93.2 \%$.

The polarized responses for $\theta = 90^o$ are shown in Fig. \ref{fig-resppoltheta90}. 
For this value of $\theta$, we observe that for $W^{TL'}_L$, the single first-order approximation and the double first-order approximation differ visibly from the
full expression  for $\eta > 0.2$. For $\eta = 0.2$, the ratio of the single first-order approximation to the full result is $96.7 \%$, and for the double first-order approximation, the ratio is $97.1 \%$.
For $\eta = 0.5$, these ratios are $ 83.5 \%$ and $85.2 \%$, respectively.

For the remaining responses, namely $W^{TL'}_S, W^{T'}_L$, and
$W^{T'}_S$, the single and double first-order approximations coincide. For the $W^{T'}$ responses, this is the case for any value of $\theta$, and for $W^{TL'}_S$, for $\theta = 90^o$ the term containing $\kappa$ is zero
due to a factor $\cos \theta$ multiplying it. 

For these three responses, the first-order approximations and the full expression agree for $\eta < 0.3$, and differences for higher values of $\eta$ are minimal. Namely, for $\eta = 0.2$, the first-order to full ratio is
$101  \%$ for $W^{TL'}_S$, $100 \%$ for $W^{T'}_L$ and $99.9 \%$ for $W^{T'}_S$. For $\eta = 0.5$, the 
first-order to full ratio is
$104  \%$ for $W^{TL'}_S$, $102  \%$ for $W^{T'}_L$ and $99.1 \%$ for $W^{T'}_S$.

The above calculations and figures simply serve to illustrate that the quality of the single first-order approximation and the double first-order approximation vary strongly with the kinematics and the response 
function considered. When considering other values of $\tau$, both smaller and larger than the value of 
$\tau = 0.25$ we used here, the results are qualitatively extremely similar. However, the results for 
assuming a 
neutron as the target, {\it i.e.,} using neutron electric and magnetic form factors in our calculations, are in a few 
cases a bit different: apart from the value for some responses, {\it e.g.,} the longitudinal response, being considerably
smaller due to the much smaller value of $G_E^n$ compared to $G_E^p$, the leading term proportional 
to $G_E$ may now be relatively smaller than the higher-order terms that are proportional to $G_M$. This leads, for some responses and in some kinematics, to a larger divergence between the full solution and either version
of first-order approximation. Similar effects may be observed when using isovector form factors, relevant for 
charge-changing neutrino scattering, where the electric term, $G_E^v$, is essentially $G_E^p\sim G_D$, while the magnetic term, $G_M^v = G_M^p - G_M^n\sim 4.70 G_D$, is significantly larger.

\section{\protect\bigskip Conclusions \label{sec-concl}}

We have presented equations for the response functions appearing when polarized leptons scatter elastically from polarized nucleons, both 
the exact results and systematic approximations involving expansions in terms of $\eta$, the dimensionless momentum of the initial nucleon. In what is called the single first-order approximation, we approximated any terms with explicit $\eta$ dependence, while retaining the distinction between $\tau$ and $\kappa$, the dimensionless 4- and 3-momentum transfers, respectively. In what is called the double first-order approximation, we used $\tau$ as an independent variable, but then expanded $\kappa$ in terms of $\eta$, up to first order in $\eta$. We have presented all of these expressions in detail, thus extending our past work \cite{JandD} for the unpolarized scattering.

In our presentation of the responses we have used the Relativistic Fermi Gas model to guide us in identifying what we dub the ``Fermi cone" and what lies outside that kinematic regime, dubbed the ``tail region". Note that nothing depends in detail on having invoked the RFG and it is only employed as a rough guide in the present work. On the one hand, in some applications one may be focused on results in the former region and there expansions in $\eta$ of the above types may or may not be appropriate --- such observations are detailed in the present study. On the other hand, when the tail region is the one of interest expansions in $\eta$ are typically counter-indicated.

In summary, we have found that, while for some responses the single fist-order approximation does a decent job, especially at lower values of $\eta$, for other responses, especially the polarized ones, there is a clear difference between the full result and the single first-order approximation once even moderate values of $\eta$ are reached. We conclude that using the double first-order approximation is probably not a good idea, and that the use of the single first-order approximation should be carefully weighed for larger $\eta$ values.

Our numerical illustrations have been performed for the proton as target. Note that for a neutron target, the current contributions involving the electric form factor
will be suppressed since $|G_E^n| \ll G_E^p$, and thus higher-order terms in $\eta$ may gain more relative weight. Also, when considering isoscalar and isovector form factors, {\it e.g.,} for neutrino scattering, one may run into situations where the weighting of the leading terms through the form factors is less pronounced, and  
higher-order terms may become more relevant. The numerical calculations for scattering from a proton target presented here are, in a way, the best case for making the discussed approximations.

Having these electromagnetic response functions for a wide range of kinematics, specifically, in the Fermi cone and tail regions, they also serve to shed some light on other traditional approaches to prescriptions for nuclear physics, in which 
approximations in $\eta$, $\kappa$ or $\lambda$ are often invoked. For instance, practical use usually demands such approximations when working with the current operators in coordinate space.  Accordingly we have included some discussion on the general form of the current operators. Clearly, the appropriateness of these approximations depends on the kinematics and
the observable considered. We conclude that, at least in the quasielastic regime, great care should be exercised when working in coordinate space where traditional approximations may be inadequate. 

\appendix

\section{Conventions\label{sec-conventions}}

In this work we employ the conventions adopted in previous work including \cite{Donnelly:1985ry,Raskin:1988kc,Donnelly:2023rej,arxivlong}: 4-vectors are written $%
A^{\mu }=(A^{0},A^{1},A^{2},A^{3})=(A^{0},\mathbf{a})$ with capital letters
for the 4-vectors and lower-case letters for 3-vectors. The magnitude of a
3-vector is written as $a=|\mathbf{a}|$. One also has $A_{\mu }=g_{\mu \nu
}A^{\mu }=(A^{0},-A^{1},-A^{2},-A^{3})$ with%
\begin{equation}
g_{\mu \nu }=g^{\mu \nu }=\left( 
\begin{array}{llll}
1 & 0 & 0 & 0 \\ 
0 & -1 & 0 & 0 \\ 
0 & 0 & -1 & 0 \\ 
0 & 0 & 0 & -1%
\end{array}%
\right) .  \label{eq-app-conv-1}
\end{equation}%
The scalar product of two 4-vectors is given by $A\cdot B=A_{\mu }B^{\mu
}=(A^{0})^{2}-a^{2}$, following the conventions of \cite{Bjorken:1965sts}. For instance,
for the 4-momentum of an on-shell particle of mass $M$, energy $E_p$ and
3-momentum $p$ we have $P^{\mu }=(E_p ,\mathbf{p})$ and hence $P^{2}=P_{\mu
}P^{\mu }=E_p^{2}-p^{2}=M^{2}$. One problem occurs with these conventions, 
\textit{viz.} for the momentum transfer 4-vector we have $%
Q^{2}=(Q^{0})^{2}-q^{2}$ which, for electron scattering is spacelike, and
accordingly $Q^{2}<0$. One should be careful not to confuse our sign convention for this quantity with the so-called SLAC convention which has the opposite sign. 

The totally anti-symmetric Levi-Civita symbol follows the conventions
of \cite{Bjorken:1965sts} where%
\begin{equation}
\epsilon _{0123}=-\epsilon ^{0123}=+1.  \label{eq-app-conv-2}
\end{equation}%
When applying the Feynman rules we also employ the conventions of \cite{Bjorken:1965sts}.

We take $\hbar = c =1$.


{\bf Acknowledgments}: This work was
supported in part by funds provided by the National Science Foundation under
grant No. PHY-2208237 (S. J.).


\end{document}